\documentclass[12pt]{article}
\usepackage{amsfonts,amssymb,graphicx,fullpage}
\usepackage[tight]{subfigure}
\usepackage{rotating}
\pdfoutput=1
\usepackage{ifpdf}
\ifpdf
  \usepackage{color}
  \definecolor{darkblue}{rgb}{0.3,0.3,0.6}
  \usepackage[pdftitle={Yukawas for the SM on Z6'},pdfauthor={Gabriele Honecker and Joris Vanhoof},pdfsubject={String theory},pdfkeywords={string theory},colorlinks=true,linkcolor=black,urlcolor=darkblue,filecolor=darkblue,citecolor=darkblue,menucolor=darkblue,breaklinks=true,bookmarks=true,anchorcolor=black,unicode=true]{hyperref}
\else
  
\fi
\usepackage[numbers,compress]{natbib}
\usepackage{a4wide}
\usepackage{longtable}
\usepackage{graphicx}
\usepackage{subfigure}
\usepackage[centertags]{amsmath}

\newcommand{\sgn}{{\rm sgn}}
\newcommand{\unity}{{\footnotesize\mbox{1\!\!I}}}

\def\muc{\multicolumn}

\def\Z{\mathbb{Z}}
\def\unity{1\!\!{\rm I}}
\def\ov{\overline}
\def\N{\mathbf{N}}
\def\Sym{\mathbf{Sym}}
\def\Anti{\mathbf{Anti}}
\def\Adj{\mathbf{Adj}}

\def\ov{\overline}
\def\1{{\bf 1}}
\def\2{{\bf 2}}
\def\3{{\bf 3}}
\def\4{{\bf 4}}
\def\6{{\bf 6}}
\def\OR{\Omega\mathcal{R}}

\def\targ#1#2{\genfrac{[}{]}{0pt}{}{#1}{#2}}
\def\tarh#1#2{\genfrac{(}{)}{0pt}{}{#1}{#2}}

\newcommand{\bCaptionfonts}{\small}
\makeatletter
\long\def\@makecaption#1#2{%
  \vskip\abovecaptionskip
  \sbox\@tempboxa{{\bCaptionfonts #1: #2}}%
  \ifdim \wd\@tempboxa >\hsize
    {\bCaptionfonts #1: #2\par}
  \else
    \hbox to\hsize{\hfil\box\@tempboxa\hfil}%
  \fi
  \vskip\belowcaptionskip}
\makeatother

\makeatletter
\let\ORIGINALlatex@openbib@code=\@openbib@code
\renewcommand{\@openbib@code}{\ORIGINALlatex@openbib@code\setlength{\itemsep}{1ex plus.5ex minus.5ex}\setlength{\parsep}{0pt}}
\makeatother

\allowdisplaybreaks[4]

\begin{document}
\begin{center}
\begin{flushright}
{\small MZ-TH/12-04\\January 2012}
\end{flushright}

\vspace{25mm}
{\Large\bf Yukawa couplings and masses of non-chiral states for the Standard Model on D6-branes on  $T^6/\Z_6'$} 

\vspace{12mm}
{\large Gabriele Honecker${}^{\heartsuit}$ and Joris Vanhoof${}^{\spadesuit,\diamondsuit,}$\footnote{Aspirant FWO.}
}

\vspace{8mm}
{~$^{\heartsuit}$\it Institut f\"ur Physik  (WA THEP), Johannes-Gutenberg-Universit\"at, D-55099 Mainz, Germany
\; {\tt Gabriele.Honecker@uni-mainz.de}}\\[1ex]
{~$^{\spadesuit}$\it Theoretische Natuurkunde, Vrije Universiteit Brussel and The International Solvay Institutes, Pleinlaan 2, B-1050 Brussels, Belgium
\; {\tt Joris.Vanhoof@vub.ac.be}}\\[1ex]
{~$^{\diamondsuit}$\it Institute of Theoretical Physics, K.U.Leuven, Celestijnenlaan 200D, B-3001 Leuven, Belgium}

\vspace{15mm}{\bf Abstract}\\[2ex]\parbox{140mm}{
The perturbative leading order open string three-point couplings for the Standard Model with hidden $USp(6)$ on fractional D6-branes on $T^6/\Z_6'$
from~\cite{Gmeiner:2008xq,Gmeiner:2009fb} are computed. Physical Yukawa couplings consisting of holomorphic Wilsonian superpotential
terms times a non-holomorphic prefactor involving the corresponding classical open string K\"ahler metrics are given,
and mass terms for all non-chiral matter states are derived. The lepton Yukawa interactions are at leading order flavour diagonal,
while the quark sector displays a more intricate pattern of mixings.
While $\mathcal{N}=2$ supersymmetric sectors acquire masses via only two D6-brane displacements - which also provide 
the hierarchies between up- and down-type Yukawas within one quark or lepton generation -, the remaining vector-like states receive 
masses via perturbative three-point couplings to some Standard Model singlet fields with {\it vev}s along flat directions.\\
Couplings to the hidden sector and messengers for supersymmetry breaking are briefly discussed.
}
\end{center}

\thispagestyle{empty}
\clearpage

\tableofcontents
\newpage
\setlength{\parskip}{1em plus1ex minus.5ex}
%%%%%%%%%%%%%%%%%%%%%%%%%%%%%%%%%%%%%%%%%%%%%%%%%%%%%%%%%%%%%%%%%%%%%%%%5
\section{Introduction}\label{S:intro}

The Standard Model gauge group and charged chiral spectrum have been obtained in a variety of globally consistent 
(RR tadpole cancelling or Bianchi identity fulfilling plus constraints from K-theory) string compactifications  over the past years, 
on the one hand on orbifolds of type IIA 
orientifolds~\cite{Uranga:2003pz,Blumenhagen:2005mu,Blumenhagen:2006ci,Dudas:2006bj,Marchesano:2007de,Lust:2007kw,Cvetic:2011vz} 
and of the heterotic string~\cite{Buchmuller:2005jr,Lebedev:2006kn,Buchmuller:2006ik,Lebedev:2008un} and 
Gepner models~\cite{Dijkstra:2004ym,Dijkstra:2004cc,Anastasopoulos:2006da}, 
on the other hand on smooth Calabi-Yau backgrounds in the framework of heterotic 
strings~\cite{Braun:2005nv,Bouchard:2005ag,Blumenhagen:2005ga,Blumenhagen:2005pm,Blumenhagen:2005zg,Blumenhagen:2006ux,Anderson:2011ns} 
and of F-theory~\cite{Weigand:2010wm}.
For all classes of perturbative models, the dimensional reduction of the classical low-energy supergravity limit serves to test 
the strength of gauge couplings and to derive the closed string moduli potential at leading order, see e.g.~\cite{Camara:2011jg,Grimm:2011dx,Kerstan:2011dy} 
in the context of D6-branes, and selection rules on the existence of Yukawa 
interactions are set-up based on charge neutralness and on intersections of cohomology groups~\cite{Braun:2006me}.

On toroidal orbifold backgrounds, conformal field theory (CFT) techniques render it possible to go a step further and compute
the perturbatively exact holomorphic gauge kinetic functions of heterotic~\cite{Derendinger:1991hq,Kaplunovsky:1995jw} and type IIA 
compactifications with intersecting 
D6-branes~\cite{Lust:2003ky,Akerblom:2007np,Blumenhagen:2007ip,Gmeiner:2009fb,Honecker:2011sm,Honecker:2011hm}.\footnote{It is also possible to 
compute non-perturbative corrections to the holomorphic gauge kinetic function from D-brane instantons, see 
e.g.~\cite{Akerblom:2007uc,Billo:2007py,Billo:2007sw,Angelantonj:2009yj}.
For D-brane instanton corrections to Yukawa interactions and other superpotential contributions see e.g.
some early works~\cite{Blumenhagen:2006xt,Ibanez:2006da,Abel:2006yk}, for an extensive list see the review~\cite{Blumenhagen:2009qh}.}
In the process of matching the CFT results with the standard supergravity expressions, the K\"ahler metrics for
open string matter fields and the bulk K\"ahler potential are obtained to lowest order, see~\cite{Akerblom:2007uc} for the 
six-torus and~\cite{Honecker:2011sm,Honecker:2011hm} for toroidal orbifolds. For the six-torus, the results agree with an 
alternative derivation by means of scattering amplitudes~\cite{Lust:2004cx,Cvetic:2003ch,Abel:2003vv}, which allow to compute the physical Yukawa couplings
in perturbation theory  as a product of the non-holomorphic K\"ahler potential and K\"ahler metric contributions times 
a holomorphic worldsheet instanton sum derived in~\cite{Cremades:2003qj,Cremades:2004wa}. Very recently, it 
was shown in~\cite{Berg:2011ij} that certain one-loop corrections to the open string K\"ahler metrics are absent
in globally consistent intersecting D6-brane models.

Phenomenological implications beyond the matter content and gauge couplings on intersecting D6-branes such as the flavour 
structure have mainly been discussed for the six-torus, see e.g.~\cite{Abel:2003yh,Kitazawa:2004nf,Higaki:2005ie,Dutta:2006bp}, 
which does, however, not admit globally consistent supersymmetric intersecting D6-brane models. 
In the present work, we take a first step at filling this gap by computing all leading holomorphic worldsheet instanton
contributions to the perturbative Yukawa couplings for the supersymmetric Standard Model with `hidden' $USp(6)_h$
on intersecting D6-branes in the $T^6/\Z_6'$
orientifold background of~\cite{Gmeiner:2008xq,Gmeiner:2009fb}. In addition, we determine all open string K\"ahler metrics
for the particular model at the classical level in extension of the partial results given in~\cite{Honecker:2011sm,Honecker:2011hm}. 
The combination of  the contributions from the smallest allowed holomorphic worldsheet instanton areas with the non-holomorphic factor
involving the K\"ahler metrics  gives an estimate on the size of physical Yukawa couplings in the low-energy supergravity theory.
Along the way, we identify possible mass terms for all vector-like matter states of the model.
Besides the well known mechanism of continuous displacements or Wilson lines, this involves the computation of 
a large number of perturbative three-point interactions in analogy to the Yukawa couplings and a discussion of vacuum expectation 
values ({\it vev}s) of scalar Standard Model singlet fields which contribute to these interactions.

{\bf Outline}\\
The paper is organised as follows:
In section~\ref{S:Yukawas_SMonT6Z6p} we briefly review the geometry of the background and the localisation of matter states
of the Standard Model with hidden $USp(6)_h$ on D6-branes on $T^6/\Z_6'$ from~\cite{Gmeiner:2008xq,Gmeiner:2009fb}.
In section~\ref{S:Yukawas+masses}, three-point couplings are discussed and the selection rules and suppression factors of 
quark and lepton Yukawa couplings of the model are presented. Yukawa hierarchies and flavour structures are briefly discussed.
In section~\ref{Ss:masses}, we discuss how the abundant 
vector-like fields in the spectrum, which do not stem from microscopic ${\cal N}=2$ supersymmetric sectors of the D6-brane set-up, 
acquire masses through three-point couplings to some Standard Model singlets with {\it vev}s along flat directions. 
Section~\ref{S:Couplings_hidden} contains
a brief discussion of couplings to the hidden sector and potential messenger fields for supersymmetry breaking.
Finally, section~\ref{S:Conclusions} contains our conclusions. The complete tables of localisations of matter states 
at pairs of D6-branes and all leading open string three-point interactions of the model with both holomorphic 
and non-holomorphic suppression factors are collected in two appendices.

%%%%%%%%%%%%%%%%%%%%%%%%%%%%%%%%%%%%%%%%%%%%%%%%%%%%%%%%%%%%%%%%%%%%%%%%5

\section{Geometry of the Standard Model on $T^6/\Z_6'$ revisited}\label{S:Yukawas_SMonT6Z6p}

We briefly review the  type IIA/$\OR$ orientifold geometry of the {\bf ABa} lattice 
and orbifold fixed points on $T^6/\Z_6'$. We introduce the D6-brane configuration 
of~\cite{Gmeiner:2008xq, Gmeiner:2009fb} which provides the Standard Model spectrum plus a hidden
$USp(6)_h$ gauge group and vector-like exotic matter states, and we discuss the different localisations of matter states
on the compact space in terms of intersection sectors $x(\theta^k y)$ of (orbifold images of) D6-branes $x,y$ and various intersection points per sector. 
The localisation on specific orbifold image D6-branes such as two right-handed quarks at 
$ac$ and one at $a(\theta^2 c)$ intersections  are required in section~\ref{S:Yukawas+masses} to state selection rules for Yukawa and 
other open string three-point couplings, and the exact positions of the intersection points are needed to determine the dominant worldsheet
contributions and estimate the size of the corresponding three-point couplings.

\subsection{Geometric setup}

\subsubsection{Compact space}

\begin{figure}[!h]
\centering
\includegraphics[width=16cm]{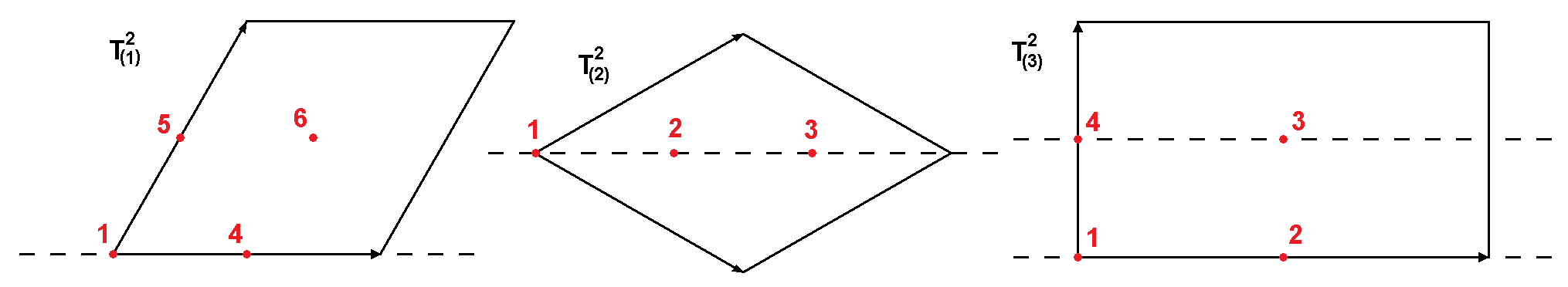}
\caption{\emph{The compact space as a direct product of three two-tori {\bf ABa} of  $T^6/\Z_6'$. The two ${\cal R}$-invariant 
O6-planes are given by the dashed horizontal lines. On the first and third torus, the marked points are invariant under the 
$\mathbb{Z}_{2}$ action, while those on the second torus are invariant under a $\Z_3$ rotation (with identical copies on the first torus,
which are not displayed).}}
\label{SixTorus}
\end{figure}

The six-dimensional compact space of the $T^6/\Z_6'$ orbifold consists of a six-torus with an additional discrete 
symmetry, see~\cite{Gmeiner:2007zz,Gmeiner:2008xq,Gmeiner:2009fb} and also~\cite{Bailin:2006zf,Bailin:2007va,Bailin:2008xx}
in the context of D6-branes in type IIA string theory.  A {\it factorisable} $T^{6}$ is separated into three different two-tori, $\otimes_{i=1}^3 T^{2}_{(i)}$,
where each of these can be represented by a parallelogram with cyclic coordinates, e.g. $(x_{1},x_{2})\sim(x_{1}+nR_{1},x_{2}+mR_{2})$ with 
$(n,m)\in\mathbb{Z}\times\mathbb{Z}$. 
The $\Z_6'$ orbifold is generated by $\theta$ acting as a rotation on the complex coordinates $z_i \equiv x_{2i-1} + i \, x_{2i}$
($i=1,2,3$),
\begin{equation*}
\theta: z_i \longrightarrow e^{2\pi i v_i} z_i
,\qquad\qquad
\text{ with }
\qquad 
\vec{v}=\frac{1}{6}(1,2,-3)
\qquad
\text{ for }
\qquad 
T^6/\Z_6'
.
\end{equation*}
The orientifold action for the type IIA string theory consists of the worldsheet parity $\Omega$ and an anti-holomorphic involution $\cal{R}$,
\begin{equation*}
{\cal R: } \, z_i \longrightarrow \ov{z}_i
, 
\end{equation*}
and it acts crystallographically on each two-torus lattice.
In figure~\ref{SixTorus}, a schematic representation is given of the three different two-tori {\bf ABa} of the model in~\cite{Gmeiner:2008xq,Gmeiner:2009fb}. 
The position of the $\OR$ invariant orientifold six-planes (O6-planes) is displayed in dashed lines. 
The orientifold symmetry $\OR$ acts as a reflection over the $\OR$ invariant O6-plane. 
The $\mathbb{Z}_{6}'$ symmetry on the six-torus can be separated into a $\mathbb{Z}_{2}$ and a $\mathbb{Z}_{3}$ part. 
The $\mathbb{Z}_{2}$ action is generated by the shift vector $3\vec{v}=\frac{1}{2}(1,0,-1)$ and acts as a point reflection over the 
origin on the first and third torus. It leaves the second torus invariant. The invariant lattice points on the first torus are denoted by (1,4,5,6) 
in figure~\ref{SixTorus} and on the third torus by (1,2,3,4). These labels are in agreement with the convention 
of~\cite{Gmeiner:2007zz}.

The remaining $\mathbb{Z}_{3}$ action is generated by the shift vector $2\vec{v}=\frac{1}{3}(1,-1,0)$
and leaves the third torus invariant.
 On the first torus the images of the $\mathbb{Z}_{2}$ invariant points under the $\Z_6'$ orbifold action are given by
\begin{equation}
T^2_{(1)}: \qquad 
\theta(1)=1, \qquad \theta(4)=5, \qquad \theta(5)=6, \qquad \theta(6)=4
.
\end{equation}
The points (1,2,3) on the second torus are invariant under the $\Z_6'$ orbifold action
since $\theta$ acts as a $\Z_3$ rotation here. On the first two-torus, however, the analogous points transform non-trivially under 
the $\Z_6'$ and $\Z_2$-subgroups,
\begin{equation}
T^2_{(1)}: \qquad 
\theta(1)=1, \qquad \theta(2)=3, \qquad \theta(3)=2
.
\end{equation}
From these data, the massless closed string spectrum can be determined, see e.g. table~9 in~\cite{Gmeiner:2007zz} and tables 57 and 58 in~\cite{Forste:2010gw}.
More details on the construction of the type IIA string
on the factorisable $T^6/\Z_6'$ orbifold and the lattice orientations may be found e.g. in~\cite{Gmeiner:2007zz}. 

In the following section, we discuss how the Standard Model with hidden $USp(6)_h$ is engineered on this particular orbifold background.

\subsubsection{D6-brane configuration}

D6-branes in our model 
are six plus one-dimensional objects along $\mathbb{R}^{1,3} \times T^6/\Z_6'$, which fill the four-dimensional Minkowski space $\mathbb{R}^{1,3}$
and wrap a three-cycle along the compact space $T^6/\Z_6'$. If we represent the six-torus as the product of three two-tori, $T^6=\otimes_{i=1}^3 T^2_{(i)}$, 
depicted in figure~\ref{SixTorus}, the three dimensions of a {\it factorisable} cycle in the six dimensional compact space are divided
into one dimension on each two-torus, i.e. a product of one-cycles.
In the example with Standard Model spectrum plus hidden sector $USp(6)_h$ of~\cite{Gmeiner:2008xq, Gmeiner:2009fb} that we consider, 
five different stacks of supersymmetric D6-branes
labelled by $a$, $b$, $c$, $d$ and $h$ satisfy the global consistency conditions of RR tadpole cancellation and K-theory constraints. 
Table~\ref{RepresentationsSMOnBranes} summarises the essential aspects of the D6-brane
configuration:
%%%%%%
\begin{table}[ht!]
\footnotesize
\begin{center}
   \begin{tabular}{|c||c|c|c|c|c|c|c|c|c|}      \hline
\multicolumn{10}{|c|}{\text{\bf D6-brane configuration for the Standard Model with hidden $USp(6)_h$ on $T^6/\Z_6'$ }}
\\ \hline\hline
      \vspace{-6mm} \begin{tabular}{c}D6-brane\\\\\\\end{tabular}
      & \begin{sideways} \hspace{2mm} label \end{sideways}
      & \begin{tabular}{c}angle on\\($T_{(1)}^2$,$T_{(2)}^2$,$T_{(3)}^2$)\\ w.r.t. $\OR$-plane\\\\\\\end{tabular}
      & \begin{sideways} \hspace{-10mm} \begin{tabular}{c}displacement $(\vec{\sigma})$\\  on ($T_{(1)}^2$,$T_{(3)}^2$)\\\end{tabular} \end{sideways}
      & \begin{sideways} \hspace{-4mm} multiplicity \end{sideways}
      &  \begin{sideways} \hspace{-8mm} \begin{tabular}{c}orientation\\vs. $\OR$-plane \end{tabular} \end{sideways}
      & \begin{tabular}{c}gauge\\group\\\\\\\end{tabular}
      & \begin{tabular}{c}wrapping\\numbers\\$(n^i,m^i)$\\\\\\\end{tabular} 
           &  \begin{sideways} \hspace{-6mm} $\Z_2$ eigenvalue \end{sideways} 
      & \begin{sideways} \hspace{-8mm} \begin{tabular}{c} Wilson line $(\vec{\tau})$\\  on ($T_{(1)}^2$,$T_{(3)}^2$)\\\end{tabular} \end{sideways}
      \\ \hline\hline
      Baryonic  & $a$ & $\left(-\pi/3,-\pi/6,\pi/2\right)$ & $(1;1)$ & $3$ &                   & $U(3)_a$     & $(1,-1;1,0;0,1)$ & $+$ & $(1,1)$      \\
      Left         & $b$ & $\left(\pi/6,-\pi/3,\pi/6\right)$   & $(1;0)$ & $2$ &                   & $U(2)_b$     & $(1,1;2,-1;1,1)$ & $-$ & $(1,0)$       \\
      Right      & $c$ & $\left(-\pi/3,\pi/3,0\right)$         & $(1;1)$ & $1$ & $\parallel$ & $USp(2)_c$ & $(1,-1;-1,2;1,0)$ & $+$ & $(1,1)$      \\
      Leptonic & $d$ & $\left(\pi/6,\pi/3,-\pi/2\right)$    & $(0;1)$ & $1$ &                  & $U(1)_d$     & $(1,1;1,-2;0,1)$ &  $+$ & $(1,1)$      \\
      Hidden   & $h$ & $\left(-\pi/3,-\pi/6,\pi/2\right)$   & $(0;0)$ & $3$ & $\perp$     & $USp(6)_h$ & $(1,-1;1,0;0,1)$ & $+$ & $(1,0)$
      \\
      \hline
   \end{tabular}
\end{center}
\caption{The five different stacks of D6-branes in the Standard Model on $T^6/\Z_6'$ with hidden $USp(6)$,
 together with their configurations on the factorised torus and the gauge groups they support.
The displacements $(\vec{\sigma})$ specify the localisation of intersection points along $T^2_{(1)} \times T^2_{(3)}$.}
\label{RepresentationsSMOnBranes}
\normalsize
\end{table}
%%%%%%%%%%%%%%%%%%%%%
it contains the angles that the D6-branes form with the horizontal ($\OR$ invariant) O6-plane. 
On $T_{(2)}^2$, there is no action of the $\mathbb{Z}_{2}$ symmetry. Therefore, the D6-branes can be continuously displaced. 
At first instance, one can take the D6-branes to pass trough the origin, i.e. point $1$ in figure~\ref{SixTorus}.
 On $T_{(1)}^2$ and $T_{(3)}^3$, the $\mathbb{Z}_{2}$ symmetry acts non-trivially, and the {\it fractional} D6-branes pass through 
$\mathbb{Z}_{2}$ invariant points. The notation $\sigma^i=0$ in the fourth column
denotes one-cycles passing through the origin (point $1$), while $\sigma^i=1$
denotes displacements along the lattice direction $\frac{\pi_{2i-1}}{2}$ (to point $4$ on $T_{(1)}^2$, point $2$ on $T_{(3)}^3$), or 
for D6-branes wrapping the one-cycle $\pi_{2i-1}$ displacements along $\frac{\pi_{2i}}{2}$ (to point $5$ on $T_{(1)}^2$, point $4$ on $T_{(3)}^2$).
The  multiplicity and thus the rank of the gauge group of each kind of D6-branes is given in the fifth column of table~\ref{RepresentationsSMOnBranes}. 
While the D6-branes $a,b,d$ wrapping generic three-cycles support unitary gauge factors, the stacks of D6-branes $c$ and $h$ require some extra attention 
as it turns out that they are their own orientifold images,
$(\theta c)'=(\theta c)$ and $(\theta h)'=(\theta h)$ (and also $c'=(\theta^{2}c)$ and $h'=(\theta^{2}h)$), where $(\theta c)$
is the rotated orbifold image of $c$ and $(\theta c)'$ its orientifold image. More specifically, 
$(\theta c)$ is parallel to the horizontal ($\OR$ invariant) O6-plane on all tori, 
and $(\theta h)$ is perpendicular to the same O6-plane along the four-torus $T^2_{(2)} \times T^2_{(3)}$ as listed in the sixth column of 
table~\ref{RepresentationsSMOnBranes}. 
Therefore, the gauge groups that these D6-branes carry are enhanced to an $USp(2)_c$ and an $USp(6)_h$ group, respectively~\cite{Gmeiner:2009fb}. 
The full gauge group of the model is thus given by $U(3)_{a}\times U(2)_{b}\times USp(2)_{c}\times U(1)_{d}\times USp(6)_{h}$ in the seventh column of
table~\ref{RepresentationsSMOnBranes}. The remaining thee columns contain the toroidal one-cycle wrapping numbers $(n^i,m^i)$ and various sign
factors needed for the construction of the complete matter spectrum along the lines reviewed in section~\ref{Ss:Strategy}.

Remember that we can displace the D6-branes on the second torus away from the origin. 
If we now move $(\theta c)$ away from the $\OR$-invariant O6-plane, 
we obtain a distinction between orbifold and orientifold image D6-branes, $c'\neq(\theta^{2}c)$. 
As a result, the gauge group $USp(2)_c$ breaks down to $U(1)_c$. 
More specifically, the fundamental representation $(\2)$ of $USp(2)_c$ splits into the representations 
$(\1)_1$ and $(\1)_{-1}$ of $U(1)_c$. Fields with $U(1)_{c}$ charge that stem from the same $USp(2)_{c}$ doublet still 
share field theoretical selection rules and formal expressions on couplings differing only in the size of the spanned 
worldsheets, since interaction terms are (up to the opposite variation of the size of worldsheets) 
invariant under the larger right-symmetric group $USp(2)_{c}$. 
The Higgs-up $H_{u}$ and Higgs-down $H_{d}$ particles originate for example from an $USp(2)_{c}$ doublet.
The same is true for the right-handed quarks, e.g. $(u_R,d_R)$, and right-handed leptons, e.g. $(e_R,\nu^e_R)$. 
Phenomenological implications of the breaking $USp(2)_c \to U(1)_c$ are discussed further in section~\ref{Sss:Breaking}.

The situation is different for the stack of D6-branes $h$ since its $\theta$-orbit is perpendicular to the $\OR\theta^{-2k}$ invariant orbit of 
O6-planes along $T^2_{(2)}$, and therefore a displacement does not alter the orientifold invariance of $h$, and $USp(6)_h$ remains unbroken.

By now, the gauge group of our model is $U(3)_{a}\times U(2)_{b}\times U(1)_{c}\times U(1)_{d}\times USp(6)_{h}$, 
where each $U(N_x)$ can be decomposed into $SU(N_x)\times U(1)_x$. 
We thus have four Abelian gauge groups, $U(1)_{a}\times U(1)_{b}\times U(1)_{c}\times U(1)_{d}$, which split into
 two anomalous massive and two anomaly-free massless Abelian gauge factors.
The generalised Green-Schwarz mechanism cancels the anomalies~\cite{Ibanez:2001nd} while giving masses of the order of the string scale 
$M_{\text{string}}$ and absorbing some axionic partners  of geometric complex structure moduli in the type IIA string theory language. 
The two independent massless anomaly-free symmetries include the hyper charge of the Standard Model~\cite{Gmeiner:2008xq,Gmeiner:2009fb},
\begin{equation*}
Q_{Y}=\frac{1}{6}Q_{a}+\frac{1}{2}Q_{c}+\frac{1}{2}Q_{d}
,
\qquad\quad \text{ and }\qquad\quad 
Q_{B-L}=\frac{1}{3}Q_{a}+Q_{d}
,
\end{equation*}
where the index $B-L$ refers to the difference between baryon and lepton number.\footnote{Note that also $Q_{c}=2Q_{Y}-Q_{B-L}$ 
is anomaly free, since it stems from the breaking of $USp(2)_c$ along some flat direction.} 
In nature,  the $B-L$ symmetry is not observed as a gauge symmetry, but rather as a global symmetry. There must thus be some mechanism 
by which these gauge bosons are rendered massive without making the hyper charge massive at the same time. There is indeed such a
candidate: the chiral multiplet of the right handed neutrino contains a scalar field, the sneutrino $\tilde{\nu}_R$, which is a singlet 
under all Standard Model gauge transformations but carries a charge under the $B-L$ symmetry. If this singlet receives
 a nonzero vacuum expectation value {\it (vev)}, it is expected to break the $B-L$ gauge symmetry. 
In the low energy limit, the $B-L$ symmetry then remains as a global symmetry.
We conclude this section by observing that the initial gauge group $U(3)_{a}\times U(2)_{b}\times USp(2)_{c}\times U(1)_{d}\times USp(6)_{h}$ of the 
D6-brane configuration is reduced to the group $SU(3)_a\times SU(2)_b\times U(1)_{Y}  \times USp(6)_h (\times U(1)_{B-L})$ 
which survives the transition to the low-energy field theory with $USp(2)_c \to U(1)_c$
broken by some {\it vev} of the complex scalar in the symmetric representation, and with $U(1)_{B-L}$ rendered massive by the 
{\it vev} of some right-handed sneutrino $\tilde{\nu}_R$. 
The remaining low-energy gauge group is nothing but the Standard Model group times an extra `hidden' $USp(6)_h$ group.

In the following section, we present the matter spectrum and the exact localisation of each massless open string state along the compact directions.

%%%%%%%%%%%%%%%%%%%%%%%%%%%%%%%%%%%%%%%%%%%%%%%%%
\subsection{Full particle spectrum}

\subsubsection{General strategy}\label{Ss:Strategy}

The computation of the massless matter spectrum is based on inspection of the vanishing or non-vanishing of some intersection angle among D6-branes $x$ and  $y$ 
on all three two-tori $T^2_{(i)}$ and on the $\Z_2$ and $\OR\theta^{k}$ transformation properties of the intersection points and the localised massless open string states. 
If on the one hand an intersection point at angles is not at a $\mathbb{Z}_{2}$ invariant position along $T^2_{(1)} \times T^2_{(3)}$, 
there exists an image point under the $\mathbb{Z}_{2}$ symmetry, and a chiral multiplet is localised on such a pair of $\Z_2$ image points. 
If on the other hand the intersection point at angles is $\mathbb{Z}_{2}$ invariant, there might or might not be a chiral multiplet depending 
on whether the massless state is projected out by the $\mathbb{Z}_{2}$ symmetry or not.
Instead of explicitly computing each massless state and determining its Chan-Paton label, the intersection numbers of the corresponding D6-branes
can be used as follows, more details on the matching of methods can be found in~\cite{Gmeiner:2008xq} and appendix A.2 of~\cite{Gmeiner:2009fb}.
As a starting point, the chirality due to the intersection of D6-branes $x$ and $y$ can be computed. 
We first calculate the intersection number on the factorisable six-torus,
\begin{equation}
I_{xy} \equiv \prod_{i=1}^{3}(n^{i}_{x}m^{i}_{y}-m^{i}_{x}n^{i}_{y}),
\end{equation}
where $(n^{i}_{x},m^{i}_{x})$ are the wrapping numbers of D6-brane $x$  along the basic one-cycles $\pi_{2i-1}$ and $\pi_{2i}$ on the $i$-th two-torus $T^2_{(i)}$. 
For an ordinary six-torus, this number suffices. However, the compact space in our model is a $\Z_{2N}$ orbifold of the six-torus. Therefore we 
also need to calculate the number $I_{xy}^{\mathbb{Z}_{2}}$ of $\mathbb{Z}_{2}$ invariant intersections among D6-branes $x$ and $y$. This number depends on the 
$\mathbb{Z}_{2}$ eigenvalue $\pm 1$ of the massless open string state at the intersection point and on the fact whether or not there is a relative Wilson line 
between the D6-branes $x$ and $y$, for more details see~\cite{Gmeiner:2007zz, Gmeiner:2008xq, Gmeiner:2009fb}. The intersection number 
of fractional D6-branes on $T^6/\Z_{2}$ and the corresponding net-chirality of bifundamental matter states is then given by
\begin{equation}\label{Eq:Def-Jxy-3angles}
\chi_{xy} \equiv - \Pi_x^{\rm frac} \circ \Pi_y^{\rm frac}=\frac{1}{2}\left(I_{xy}+I_{xy}^{\mathbb{Z}_{2}}\right)
.
\end{equation}
The absolute value $\varphi_{xy}^{\text{3 angles}} \equiv |\chi_{xy}|$ gives the total number of chiral multiplets at the intersection of D6-branes $x$ and $y$,
and the sign contains information on the orientation of the open strings and therefore on the chirality (or representation) of the corresponding multiplet. 
The total numbers  $\sum_{k=0}^2 \chi_{x(\theta^k y)},\sum_{k=0}^2 \varphi_{x(\theta^k y)}$ including the three $\Z_3$ orbifold images on $T^6/\Z_6'$
for the Standard Model with hidden $USp(6)_h$ were calculated in~\cite{Gmeiner:2007zz, Gmeiner:2008xq, Gmeiner:2009fb},
while in  appendix~\ref{App:A} of this article we display the full set of individual contributions $\chi_{x(\theta^k y)}, \varphi_{x(\theta^k y)}$ with $k=0,1,2$ 
here for the first time. Also the localisation of each massless matter state at intersection points on $T^6/\Z_6'$ is presented here for the first time.
If $\chi_{xy}>0$, we use the convention that open strings are oriented from D6-brane $y$ to $x$ (which will be denoted as $[xy]$),  and the associated chiral 
multiplets carry the representation $(\ov{\N}_{x},\N_{y})$, where $\N_{x}$ is the fundamental representation of the gauge group $U(N_x)$ of D6-brane stack $x$, 
while $\ov{\N}_{x}$ is the conjugate representation. If $\chi_{xy}<0$, open strings are oriented from D6-brane $x$ to $y$ ($[yx]$) and the matter fields 
transform in the representation $(\N_{x},\ov{\N}_{y})$. If $x'$ is the orientifold image of some D6-brane $x$, then $\N_{x'}=\ov{\N}_{x}$.

For D6-branes $x$ and $y$ coincident along the torus $T^2_{(2)}$ with trivial $\Z_2$ action, the charged spectrum  consists of 
${\cal N}=2$ supersymmetric non-chiral pairs of multiplets transforming as $(\N_x,\ov{\N}_y) + (\ov{\N}_x,\N_y)$ counted by the number 
of intersections~\cite{Gmeiner:2008xq, Gmeiner:2009fb}
\begin{equation}\label{Eq:Def-Jxy-angle2=0}
\varphi_{xy}^{|| \text{ on } T^2_{(2)}} \equiv \left| I_{xy}^{(1\cdot 3)}+I_{xy}^{\mathbb{Z}_{2},(1\cdot 3)} \right|
\qquad\text{ along } \qquad
T^2_{(1)} \times T^2_{(3)}
,
\end{equation}
where again $I_{xy}^{\mathbb{Z}_{2},(1\cdot 3)}$ contains sign factors from relative $\Z_2$ eigenvalues and discrete Wilson lines.
If the D6-branes coincide either on $T_{(1)}^2$ or $T_{(3)}^2$, the net-chirality~(\ref{Eq:Def-Jxy-3angles}) is non-vanishing. 
However, since there exist two states with opposite chirality and opposite $\Z_2$ eigenvalue in this sector, additional non-chiral matter pairs arise
if the intersection points along $T^2_{(2)} \times T^2_{(3)}$ or $T^2_{(1)} \times T^2_{(2)}$ form $\Z_2$ pairs. The total number 
of representations transforming as $(\N_x,\ov{\N}_y)$ or $(\ov{\N}_x,\N_y)$ is given by~\cite{Gmeiner:2008xq,Gmeiner:2009fb}
\begin{equation}\label{Eq:Def-Jxy-angle1or3=0}
\varphi_{xy}^{|| \text{ on } T^2_{(1)}} \equiv \left| I_{xy}^{(2\cdot 3)}\right|
,
\qquad\qquad
\varphi_{xy}^{|| \text{ on } T^2_{(3)}} \equiv  \left| I_{xy}^{(1\cdot 2)}\right|
. 
\end{equation}
If the D6-branes are parallel but not coincident on some two-torus $T^2_{(i)}$, the matter spectrum is non-chiral and 
massive with the mass proportional to the distance of the D6-branes,
 \begin{equation}\label{Eq:M-parallel}
M^2_{xy} \sim d^2_{(i)}(x,y)
.
\end{equation}
In the special case where the D6-brane $y$ is the same as $x$ or some $\Z_6'$ orbifold image ($\theta^{k}x$), the representation becomes 
$(\N_{x},\ov{\N}_{x})=(\Adj_x)$, i.e. the adjoint representation of the gauge group $U(N_x)$ of D6-brane $x$. 
Similarly, if the D6-brane $y$ is some $\theta$-image of the orientifold image $(\theta^{k}x')$, the representation becomes 
$(\N_{x},\ov{\N}_{x'})=(\N_{x},\N_{x})$. This representation is reducible and decomposes into a symmetric $(\Sym_x)$ and an antisymmetric 
$(\Anti_x)$ representation. 
The existence of matter in one or both representations depends on the orientifold transformation properties of the 
intersection points as well as the $\OR$ eigenvalue of the associated massless open string state, for a detailed discussion 
see~\cite{Gmeiner:2007zz,Gmeiner:2008xq} and for the complementary calculation of beta function coefficients via gauge threshold 
amplitudes see~\cite{Gmeiner:2009fb,Honecker:2011sm,Honecker:2011hm}.

The discussion of this section is general for D6-branes on orbifold in two ways: the existence of fractional D6-branes stuck at
fixed points of the $\Z_2$ symmetry on $T^6/\Z_{2N}$ and the corresponding formulas~(\ref{Eq:Def-Jxy-3angles}) and~(\ref{Eq:Def-Jxy-angle1or3=0})
also apply to the $T^6/\Z_4$ orbifold of~\cite{Blumenhagen:2002gw} and the $T^6/\Z_6$ models in~\cite{Honecker:2004kb,Honecker:2004np,Gmeiner:2007we}.
In addition, D6-branes $x$ have orbifold images $(\theta^k x)$ on any $T^6/\Z_{2N}$ or $T^6/\Z_2 \times \Z_{2M}$ background with $M \neq 0$,
see e.g.~\cite{Honecker:2003vq,Honecker:2004np,Cvetic:2006by} for models on $T^6/\Z_2 \times \Z_4$ and~\cite{Forste:2000hx,Forste:2010gw} for other 
$T^6/\Z_N \times \Z_M$ orbifold 
backgrounds. 

By the above described method, not only the complete, chiral plus vector-like, massless matter content can be derived, but also 
the localisation of each multiplet on $T^6/\Z_6'$ can be determined, which is a necessary prerequisite for computing the size of Yukawa-like
three-point interactions due to worldsheet instantons in section~\ref{S:Yukawas+masses}.

%%%%%%%%%%%%%%%%%%%%%%%%%%%%%%%%

\subsubsection{The Standard Model on $T^6/\Z_6'$: massless states and localisations}

\begin{figure}[!h]
\centering
\includegraphics[width=16cm]{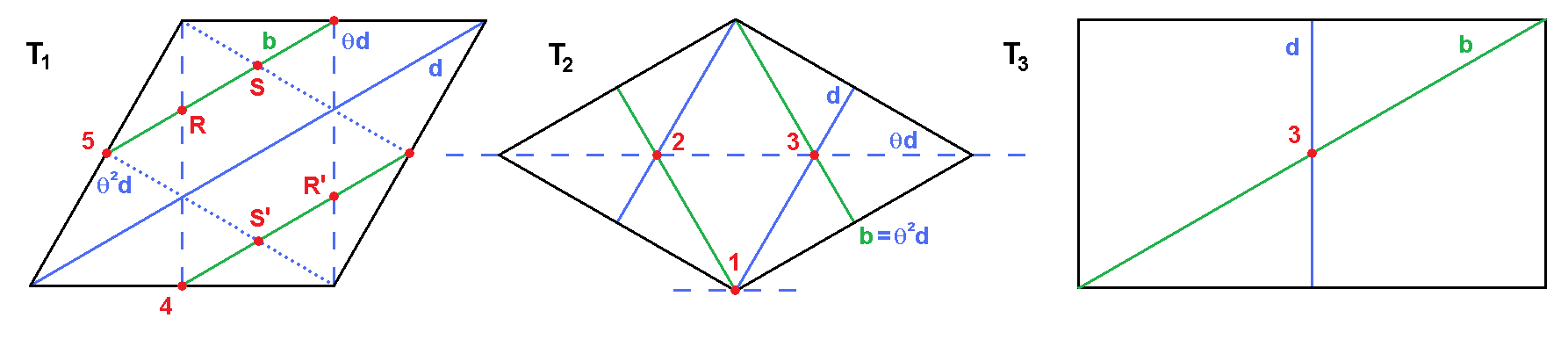}
\caption{\emph{Configuration of the stack of D6-branes $b$ (green) and D6-brane $d$ (blue) with its orbifold images $(\theta^k d)_{k=1,2}$ 
(dashed and dotted blue) on the factorisable six-torus.
Points 1 \ldots 6 correspond to $\Z_2$ or $\Z_3$ orbifold fixed points, while $R\stackrel{\Z_2}{\leftrightarrow}R'$, $S\stackrel{\Z_2}{\leftrightarrow} S'$
  correspond to points not fixed under the orbifold symmetry.}}
\label{DBIntersections}
\end{figure}

Let us explain the counting and localisation of matter states by the explicit example of the intersection points between the stack of D6-branes $b$ and 
the orbit of $\Z_6'$ images $(\theta^k d)_{k=0\ldots 2}$ of D6-brane $d$, at which the left-handed leptons $L_i$ are localised. 
The three-cycles wrapped by these D6-branes are depicted in figure~\ref{DBIntersections}, 
and labels for the intersection points are given, see also  table~\ref{IntersectionPointsBD}.

\begin{table}[ht!]
\begin{center}
   \begin{tabular}{|c|c|c|c|c|c|}      \hline
\multicolumn{6}{|c|}{\text{\bf Matter multiplicities and localisations of leptons}}
\\ \hline\hline
 Particle &      $xy$           &  $\chi_{xy}$  or $\varphi_{xy}$   & $T_{(1)}^2$         & $T_{(2)}^2$       & $T_{(3)}^2$ \\ \hline\hline
 &     $bd$           & $\emptyset$ & $\parallel\neq$ & $1$, $2$, $3$ & $3$     \\
 $L_{i, i =1 \ldots 6}$  &   $b(\theta d)$    & $-6$        & $4$, $(R,R')$   & $1$, $2$, $3$ & $3$     \\
   &   $b(\theta^{2}d)$ & $|4|_{m}$   & $5$, $(S,S')$   & $\parallel=$  & $3$     \\
      \hline
   \end{tabular}
\end{center}
\caption{Example: the total matter content $\varphi_{xy}$ for non-chiral sectors $|\cdot|$ or net-chirality $\chi_{xy}$ with sign for sectors with 
$\varphi_{xy}=|\chi_{xy}|$ at the intersection points between the stacks of D6-branes $b$ and $(\theta^k d)_{k=0\ldots 2}$ per two-torus.
The index $m$ indicates that the corresponding states stem from ${\cal N}=2$ sectors and acquire a mass upon relatively displacing the D6-brane 
stacks along $T^2_{(2)}$ in a continuous way.
$\parallel\neq$ denotes parallel displaced D6-branes supporting massive string states only, and $\parallel=$  corresponds to coincident D6-brane positions
which can support massless string states. 
The labels of intersection points are defined in figure~\protect\ref{DBIntersections}. The complete list of massless matter localisations
and their K\"ahler metrics for the Standard Model with hidden $USp(6)_h$ on $T^6/\Z_6'$ is given in table~\protect\ref{IntersectionNumbers-1} 
of appendix~\protect\ref{App:A} for bifundamental and table~\protect\ref{IntersectionNumbers-2} for symmetric and antisymmetric matter.}
\label{IntersectionPointsBD}
\end{table}
%%%%%
The interpretation is as follows. The D6-branes $b$ and $d$ do not intersect on the first torus, instead they are parallel but not coincident and do not 
support any massless matter state in the bifundamental representation. This conclusion agrees with $\chi_{bd}=\emptyset$ on the first line of 
table~\ref{IntersectionPointsBD}.
The D6-branes $b$ and $(\theta d)$ intersect once on $T_{(3)}^2$, while on $T_{(2)}^2$ they intersect in three different points. On $T_{(1)}^2$, the 
D6-branes intersect in the three points $4, R, R'$, where $R$ and $R'$ are each others' image under the $\mathbb{Z}_{2}$ symmetry. Since the 
D6-branes $b$ and $(\theta d)$ are at angles on all three two-tori, there is one chiral multiplet localised at the points $R$ and $R'$ together. 
In other words, on $T_{(1)}^2$ there exist two $\Z_2$ invariant orbits of intersection points localised at point 4
and the other spread over $R$ and $R'$. Therefore, in total,  six $\Z_2$ invariant orbits of  intersection points exist on the six-torus. 
This number matches the number $|\chi_{b(\theta d)}|=6$, and none of the massless states is projected out.
Finally, for the D6-branes $b$ and $(\theta^{2}d)$, there exist one intersection point on $T_{(3)}^2$ and two on $T_{(1)}^2$. 
However, on $T_{(2)}^2$ the D6-branes are parallel. If we continuously displace one of the D6-brane stacks on the second torus, 
they are still parallel but no longer coincident. The mass of the strings stretching between $b$  and $(\theta^{2}d)$
 scales with the distance between the D6-branes along $T^2_{(2)}$, cf. equation~(\ref{Eq:M-parallel}). 
The corresponding bifundamental multiplets organise themselves in massive 
${\cal N}=2$ supersymmetric non-chiral matter pairs $(\1,\2)_{-1/2} + (\1,\ov{\2})_{1/2}$ of $SU(3)_a \times SU(2)_b \times U(1)_Y$. 
There exist two such pairs counted by $\varphi_{b(\theta^2 d)}=|4|_{m}$, where the index $m$ signals the option of switching on a mass term
by a continuous relative Wilson line or displacement on $T^2_{(2)}$ along a flat direction in moduli space.

In a similar manner,  all massless matter fields of the Standard Model with hidden $USp(6)_h$ on $T^6/\Z_6'$ have been localised 
at the intersection points of D6-branes on the compact space. 
The affiliations to intersection sectors are all summarised in appendix \ref{App:A} in table~\ref{IntersectionNumbers-1} for bifundamental 
and adjoint representations and table~\ref{IntersectionNumbers-2} for 
symmetric and antisymmetric matter. The precise location of the representations can be determined along the lines described above or
extracted from tables~\ref{YukawaForLeptons} to~\ref{Hidden-Sequences} 
(see also the explicit localisations in tables~\ref{IntersectionPointsBD},~\ref{tab:local_Sym_U2} and~\ref{tab:local_Adj_U2} for the left-handed leptons, 
(anti)symmetrics and adjoints of $U(2)_b$, respectively, and~\cite{Honecker:2012fn} for the remaining Standard Model particles $Q_i$, $U_i$, $E_i$,
$H_i$ and $\ov{L}_i$)
and will be important to determine the selection rules and suppression factors of the perturbatively
allowed Yukawa-like three-point interactions in section~\ref{S:Yukawas+masses}.

\subsubsection{Full spectrum of the Standard Model with hidden $USp(6)_h$}\label{The full spectrum}

In~\cite{Gmeiner:2008xq,Gmeiner:2009fb}, the full massless spectrum is derived, based on the calculation of the intersection numbers $\chi_{x(\theta^k y)}$
and the total amount of matter states $\varphi_{x(\theta^k y)}$. 
 Under the Standard Model gauge group \mbox{$SU(3)_a\times SU(2)_b\times U(1)_{Y}$}, the charges are given as $({\bf R}_{SU(3)},{\bf R}_{SU(2)})_{Q_{Y}}$ below. 
The `chiral' spectrum reads
\begin{eqnarray}
[C]&=&3\times\left[(\3,\2)_{1/6}+(\ov{\3},\1)_{1/3}+(\ov{\3},\1)_{-2/3}+(\1,\1)_{1}+(\1,\1)_{0}+2\times(\1,\2)_{-1/2}+(\1,\2)_{1/2}\right] \nonumber \\
&& +9\times\left[(\1,\ov{\2})_{-1/2}+(\1,\ov{\2})_{1/2}\right] \nonumber\\
&\equiv&3\times\left[Q_{L}+d_{R}+u_{R}+e_{R}+\nu_{R}+2\times L+\ov{L}\right]+9\times\left[H_{d}+H_{u}\right] 
, \label{FullSpectrum1} 
\end{eqnarray}
where the Higgs pairs $(H_u,H_d)$ and the lepton-anti-lepton pairs $(L,\ov{L})$ are non-chiral with respect to the Standard Model gauge group,
but arise from non-vanishing intersection numbers due to the anomalous $U(1)_b \subset U(2)_b$ symmetry of the underlying D6-brane configuration.

The `non-chiral' spectrum, 
\begin{eqnarray}
[V]&=&2\times({\bf 8}_{\Adj},\1)_{0}+10\times(\1,\3_{\Adj})_{0}+26\times(\1,\1)_{0} \nonumber \\
&& + \Biggl[(\3,\2)_{1/6}+3\times(\ov{\3},\1)_{1/3}  +3\times(\ov{\3},\1)_{-2/3} + 3\times(\ov{\3}_{\Anti},\1)_{1/3} +6\times(\1,\3_{\Sym})_{0}\nonumber \\
&& \qquad +1_{m}\times(\1,\ov{\2})_{-1/2}+1_{m}\times(\1,\ov{\2})_{1/2}+2_{m}\times(\1,\2)_{-1/2}+1_{m}\times(\1,\2)_{1/2} \nonumber \\
&& \qquad +4_{m}\times(\1,\1_{\Anti})_{0}+2_{m}\times(\1,\1)_{1}+1_{m}\times(\1,\1)_{0}+c.c.\Biggr] \label{FullSpectrum2}
,
\end{eqnarray}
consists of the (microscopically ${\cal N}=2$ supersymmetric) vector-like matter states on the last two lines, which acquire masses by continuous 
parallel relative displacements of D6-branes indexed by $m$ at their multiplicities, 
and of the (pairs of microscopically  ${\cal N}=1$ supersymmetric) 
vector-like matter states on the first two lines, which can receive masses via Yukawa-like three-point couplings as discussed in detail in section~\ref{Ss:masses}.

The above parts of the spectrum are uncharged under the `hidden' gauge group $USp(6)_h$. The final part of the massless 
open string spectrum consists of the `hidden' sector 
fields that transform non-trivially under $USp(6)_h$ with representations denoted as \mbox{$({\bf R}_{SU(3)},{\bf R}_{SU(2)};{\bf R}_{USp(6)})_{Q_{Y}}$},
\begin{eqnarray}\hspace{-2mm}
[H]&=&2\times(\1,\1;{\bf 15}_{\Anti})_{0}+(\1,\2;{\bf 6})_0 +(\1,\ov{\2};{\bf 6})_0 
+2\times\Bigl[(\1,\1;{\bf 6})_{1/2}+(\1,\1;{\bf 6})_{-1/2}\Bigr] \nonumber\\
&\equiv &\left[X_{1}+X_{2}\right]+\Pi^{+}+\Pi^{-}+ \sum_{i=1}^2 \left[ \Omega^{+}_{i}+\Omega^{-}_{i} \right]
, 
\label{FullSpectrum3} 
\end{eqnarray}
where for later convenience of shortening the notation of three-point interactions 
the representations are parameterised by $X_{i,i=1,2}$, $\Pi^{\pm}$ and $\Omega^{\pm}_{i=1,2}$ on the last line.

The states in equations (\ref{FullSpectrum1}),~(\ref{FullSpectrum2}) and~(\ref{FullSpectrum3}) constitute the complete  massless particle spectrum of the $T^6/\Z_6'$ 
Standard Model with hidden sector $USp(6)_h$. Apart from the three quark and lepton generations and some Higgses, it contains 
an abundance of vector-like fields transforming as non-chiral quark or lepton pairs or in exotic representations. 
We discuss in section~\ref{Ss:masses} below how all of these fields can obtain masses via three-point couplings with some Standard Model-singlet fields
that can acquire {\it vev}s along some flat direction, similar to the Higgs field in the Standard Model Yukawa couplings.

%%%%%%%%%%%%%%%%%%%%%%%%%%%%%%%%%%%%%%%%%%%%%%%%%%%%%%%%%%%%%%%%%%%%%%%%%%%%%%%%%%%%%%%%%%

\section{Yukawa couplings}\label{S:Yukawas+masses}

\subsection{Superpotential terms}\label{Ss:Superpotential}

Closed triangles between D6-branes result in interactions between matter fields at the apexes due to the existence of worldsheet instantons 
sweeping the enclosed areas as thoroughly explained in~\cite{Cremades:2003qj,Cremades:2004wa}, see also~\cite{Cvetic:2003ch,Abel:2003vv,Abel:2003yx}, 
and briefly reviewed here.

Suppose that three intersecting D6-branes $x$, $y$ and $z$ form a triangle on each two-torus
and that each of the intersection points carries a massless string that represents a chiral matter field $\phi_{xy}^i$, $\phi_{yz}^j$ and $\phi_{zx}^k$ in
the respective representation $(\ov{\N}_x,\N_y)$, $(\ov{\N}_y,\N_z)$ and $(\ov{\N}_z,\N_x)$ of $U(N_x) \times U(N_y) \times U(N_z)$. This configuration 
gives rise to a three-point Yukawa coupling term in the superpotential,
\begin{equation}
W= W_{ijk} \; \phi_{xy}^i \phi_{yz}^j \phi_{zx}^k
.
\label{ThreePointInteraction}
\end{equation}
If we take into account the orientations of the strings, i.e. if we have a sequence $[xy]+[yz]+[zx]$, we are guaranteed that such a term is gauge invariant 
because the fields are in the  bifundamental representations of the full gauge group as stated above, and the gauge transformations cancel pair-wise in the 
expression~(\ref{ThreePointInteraction}).

A purely field theoretical analysis of the matter representations would lead to allowing additional gauge invariant terms in the superpotential that are 
actually forbidden by the D6-brane model because there exists no closed triangle. 
Therefore, the D6-brane setup in type IIA string theory serves as a selection rule on allowed interaction terms once the localisation of each matter state
in a given model is known. 
This is a strong restriction and goes well beyond the argument of gauge invariance as we will show below for 
several excluded types of couplings in the Standard Model spectrum with hidden $USp(6)_h$
on fractional D6-branes in the $T^6/\Z_6'$ background of~\cite{Gmeiner:2008xq,Gmeiner:2009fb}.

We can extract even more information from the D6-brane model. The size of the triangle formed by three D6-branes is related to the strength of the 
coupling~\cite{Cremades:2003qj,Cremades:2004wa,Cvetic:2003ch,Abel:2003vv,Abel:2003yx} as follows. Denoting the area of the enclosed triangle by $\mathcal{A}_{ijk}$, 
which we will often give as a fraction of the total areas $\mathcal{A}_{(i)}$ of the two-tori $T^{2}_{(i)}$, the coefficient of the three-point interaction 
term~(\ref{ThreePointInteraction}) is proportional to
\begin{equation}
W_{ijk} \sim e^{-\mathcal{A}_{ijk}/(2\pi\alpha')}
.
\end{equation}
Here $\tau_s=1/(2\pi\alpha')$ is the string tension. If the three D6-branes intersect exactly in one point, the area $\mathcal{A}_{ijk}$ is zero and the coefficient 
is of order $\mathcal{O}(1)$. The larger the area is, the smaller the coefficient and the stronger the suppression of the interaction. 
While the superpotential receives contributions from all possible worldsheet instantons by summing over all
images on the covering space of the torus~\cite{Cremades:2003qj,Cremades:2004wa,Cvetic:2003ch,Abel:2003vv,Abel:2003yx}
leading to a formula of the type~(\ref{Eq:Yukawas_complete_bcthetad}) stated below,
in this article we focus only on the leading order and estimate the size of the holomorphic Yukawa couplings $W_{ijk}$ by the smallest possible triangles.

The physical size of the quark- and lepton-Yukawa couplings is given by the product of the holomorphic superpotential coupling
$W_{ijk}$ in equation~(\ref{ThreePointInteraction}), which arises classically, times non-holomorphic quantum contributions involving the 
K\"ahler metrics of the corresponding matter fields~\cite{Cremmer:1982en,Dixon:1989fj,Cvetic:2003ch,Abel:2003yx,Lust:2004cx} as briefly reviewed 
in the context of the present D6-brane model in section~\ref{Ss:PhysicalYukawas}.
A second potential source of Yukawa hierarchies beyond the size of triangular instantonic worldsheets is therefore given by
the different expressions for the K\"ahler metrics.

At this point, it should be noted that additional symmetries of the string compactification, such as the $\Z_2$ eigenvalues of 
the massless states, might lead to further selection rules, which can only be determined by more sophisticated methods such
as an explicit computation of three-point couplings on the type IIA orientifold on $T^6/\Z_{2N}$ using vertex operators in generalisation of the 
recently computed two-point correlators on $T^6/\Z_6'$ in appendix C of~\cite{Berg:2011ij}.

%%%%%%%%%%%%%%%%%%%%%%%%%%%%%%%%%%%%%%%%%%%%%%%%
\subsection{Yukawa couplings for the Standard Model on $T^6/\Z_6'$}\label{Ss:Yukawa-SM}

As an example, we consider the Yukawa couplings between the left- and right handed leptons and the Higgs particles
in  the Standard Model spectrum with hidden $USp(6)_h$ on $T^6/\Z_6'$~\cite{Gmeiner:2008xq,Gmeiner:2009fb}.
For the sake of simplicity of the discussion we present here the D6-brane configuration with the right-symmetric group $USp(2)_c$ and the
corresponding right-handed lepton doublets  $E_i \equiv (e_R^i,\nu^i_R)$ and Higgs-doublets $H_i \equiv (H_u^i,H_d^i)$. 
In section~\ref{Sss:Breaking}, we briefly comment on the changes in size of the areas $\mathcal{A}_{ijk}$ swept by worldsheet instantons when 
the right-symmetric group is broken, $USp(2)_c \to U(1)_c$, by displacing the D6-brane $c$ away from the $\OR$ invariant position along the two-torus $T^2_{(2)}$
without $\Z_2$ action, and when other D6-brane displacements are applied as well to render matter in microscopically ${\cal N}=2$ supersymmetric sectors 
in the last two lines of equation~(\ref{FullSpectrum2}) massive, cf.~\cite{Gmeiner:2008xq,Gmeiner:2009fb}.
The left handed leptons $L_i$, $L_{3+i}$ with $i\in\{1,2,3\}$ labelling the localisation on $T^2_{(2)}$ 
arise at intersections of the D6-branes $b$ and $(\theta d)$ as detailed in table~\ref{IntersectionPointsBD}, while the right handed 
leptons $E_i$ localised at points $i\in \{1,2,3\}$ on $T^2_{(2)}$ stem from the intersections of $c$ with $(\theta d)$, see table~\ref{IntersectionNumbers-1} for a complete list of all
bifundamental and adjoint matter allocations to the intersection sectors. 
The Higgs generations $H_{i}$, $H_{3+i}$, $H_{6+i}$ with $i\in \{1,2,3\}$ again labelling the localisation on $T^2_{(2)}$
are partially located at intersections of the D6-brane stack $b$ with $c$ (for $H_i$, $H_{3+i}$) and partly with $(\theta c)$ (for $H_{6+i}$). 
Since allowed couplings originate from closed triangles between D6-branes, we can immediately deduce that the latter three kinds of Higgs fields $H_{6+i}$ do 
not couple directly to the leptons. This is because the only existing closed sequence of 
the left, right and leptonic stacks of D6-branes is given by $[c(\theta d)]+[(\theta d)b]+[bc]$, thus ruling out the $[b(\theta c)]$ intersections.
\begin{figure}[!h]
\centering
\includegraphics[width=16cm]{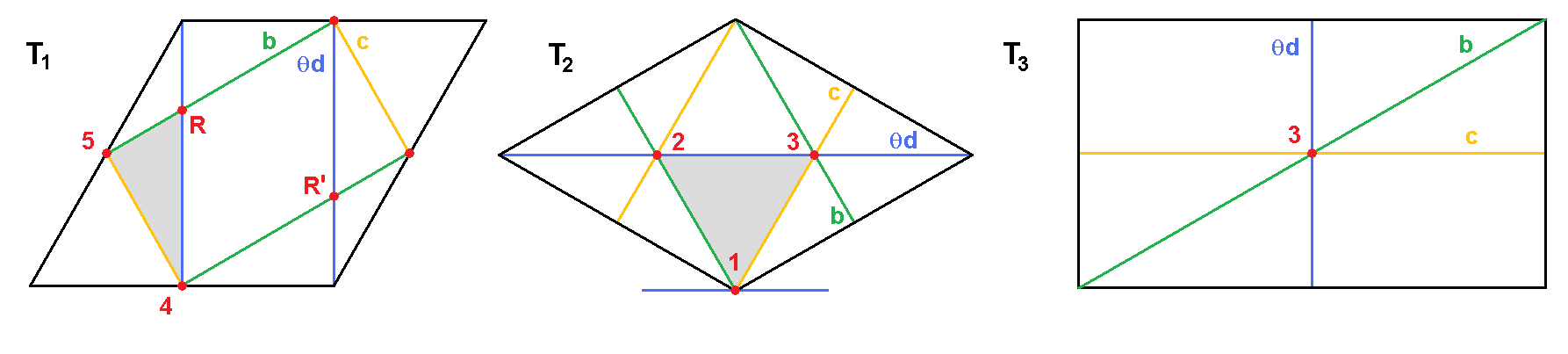}
\caption{\emph{Position of the D6-branes $b$, $c$ and $(\theta d)$ and their intersection points. The shaded areas 
correspond to examples of instantonic worldsheets generating Yukawa interactions with flavour mixing.}}
\label{LeptonsOnBranes}
\end{figure}
The positions of the D6-brane stacks $b$, $c$ and $(\theta d)$ and their intersections on the factorised six-torus are displayed in figure~\ref{LeptonsOnBranes}. 
On the third torus $T^2_{(3)}$, all three D6-branes intersect once in the same point 3. On $T_{(1)}^2$, the intersection point 4 is common to the three different 
D6-branes. Likewise,  on $T_{(2)}^2$, all  three D6-branes intersect in point 1. Therefore, we expect an interaction term between the fields that are localised in 
point (413) along $T^2_{(1)} \times T^2_{(2)} \times T^2_{(3)}$. Since the size of the corresponding enclosed triangle is zero on each two-torus, this interaction 
term is not area-suppressed. 
Inspection of the localisation of the matter fields in table~\ref{YukawaForLeptons} of appendix~\ref{App:A} identifies this interaction as that of the first
lepton with the first Higgs generation,
\begin{equation}
W_{E_{1}L_{1}H_{1}} \sim {\cal O}(1). 
\end{equation}
Let us now consider different points on the second two-torus $T^2_{(2)}$. Instead of all fields in point 1, we choose the intersection of $(\theta d)$ 
with $b$ in point 2 (left-handed lepton $L_{2}$), the intersection of $c$ with $(\theta d)$ in point 1 (right-handed lepton pair $E_{1}$) and the field 
at the intersection of $b$ with $c$ in point 3 (Higgs pair $H_{3}$). These points form the closed shaded triangle on the second two-torus in
figure~\ref{LeptonsOnBranes} and thus result in an allowed coupling, which is suppressed compared to the previously discussed flavour diagonal one, $E_1L_1H_1$. 
By simple geometric arguments one can deduce the ratio of the area of the enclosed triangle to be ${\cal A}_{(2)}/6$, where ${\cal A}_{(2)}$ is
the total dimensionless area of $T_{(2)}^2$.\footnote{In the following, all two-torus areas are measured in units of $2\pi\alpha'$.
To be in the geometric regime of the compactification, we require $\mathcal{A}_{(i)} > 1$ for $i=1,2,3$.
In slight abuse of the notation, we also use the symbol $v_i \simeq {\cal A}_{(i)}$ for the bulk K\"ahler moduli dependence of the K\"ahler metrics 
in section~\ref{Ss:PhysicalYukawas} and the last column of each table containing suppression factors of thee-point interactions.
Last but not least, $\prod_{i=1}^3 {\cal A}_{(i)} \sim \prod_{i=1}^3 v_i$ denotes the volume of the $T^6/\Z_6'$ {\it orbifold}, and we absorbed the 
numerical factors compared to the ordinary six-torus into the definition of the two-torus volumes.} 
Therefore we find the interaction term
\begin{equation}
W_{E_{1}L_{2}H_{3}} \sim {\cal O}(e^{-\mathcal{A}_{(2)}/6}).
\end{equation}
Similarly, examining the triple intersection points 1 on $T_{(2)}^2$ and 3 on $T_{(3)}^2$, but  different intersection points 4,5,$R$
on $T_{(1)}^2$ corresponds respectively the localisations of fields $E_{1}$, $H_{4}$ and $L_{4}$. Their three-point interaction is also allowed
by the existence of a closed triangle,  and the ratio of the latter to the area of $T_{(1)}^2$ as displayed in figure~\ref{LeptonsOnBranes} 
is computed as $1/12$, 
\begin{equation}
W_{E_{1}L_{4}H_{4}} \sim {\cal O}(e^{-\mathcal{A}_{(1)}/12}).
\end{equation}
Finally, there exist instantonic worldsheets due to  nonzero triangles on both $T_{(1)}^2$ and $T_{(2)}^2$.
The example depicted in figure~\ref{LeptonsOnBranes} with shaded areas on $T_{(1)}^2 \times T_{(2)}^2$
corresponds e.g. to the Yukawa interaction
\begin{equation}
W_{E_{1}L_{5}H_{6}} \sim  {\cal O}(e^{-\mathcal{A}_{(1)}/12} \cdot e^{-\mathcal{A}_{(2)}/6})=
 {\cal O}(e^{-\mathcal{A}_{(1)}/12 - \mathcal{A}_{(2)}/6})
\end{equation}
according to the labels of particle generations in the appendices.
As for the examples in tables~\ref{IntersectionPointsBD},~\ref{tab:local_Sym_U2} and~\ref{tab:local_Adj_U2}, particle generations are labelled by their localisations on $T^2_{(2)}$ 
whenever possible.

A suppression by non-vanishing triangles along the whole six-torus does not occur at leading order for the lepton and quark Yukawas 
in the Standard Model example with hidden $USp(6)_h$ on $T^6/\Z_6'$ since the D6-branes $a,b,c,d$ pair-wise intersect once  or are parallel on $T^2_{(3)}$. 
The only exception is the double intersection of stack $b$ with its orientifold image $b'$ along $T^2_{(3)}$, which leads to three-point interactions in table~\ref{LLbarSym}
suppressed by ${\cal A}_{(3)}/4$ of vector-like lepton or Higgs pairs with the symmetrics of $U(2)_b$.
At sub-leading orders, i.e. performing the worldsheet instanton sum~(\ref{Eq:Yukawas_complete_bcthetad})
over all possible copies of the defining domain of the two-tori, all three kinds of area suppressions 
by ${\cal A}_{(i)}$ with $i \in \{1,2,3\}$  will occur.

A list of all leading order lepton Yukawa couplings is given in table \ref{YukawaForLeptons} of appendix \ref{App:B}, 
and the leading quark Yukawa interactions are summarised in table~\ref{YukawaForBaryons}. The appendix also lists all allowed 
three-point interactions rendering the vector-like matter massive as discussed in section~\ref{Ss:masses}.

%%%%%%%%%%%%%%%%%%%%%%%%%%%%%%%%%%

\subsection{Remarks on phenomenological aspects}

In the previous section, we provided a detailed discussion of the suppression of holomorphic Yukawa couplings due to triangular worldsheets for 
some examples of leptonic couplings in the Standard Model example on $T^6/\Z_6'$. Besides only having considered the leading worldsheet contributions,
generically higher order and D-instanton interactions exist. The suppression factors change upon breaking the right-symmetric group by a displacement $\sigma^2_c$,
 $USp(2)_c \to U(1)_c$, and the existence of several Higgs fields leads to a complicated interaction pattern in the lepton and quark sectors.

\subsubsection{Higher $n$-point interactions} 

Renormalisability of the supersymmetric four-dimensional field theory constrains the superpotential to maximal degree three,
which corresponds to the triangular worldsheets spread among three D6-branes as discussed above for some examples of lepton couplings. 
An example for a renormalisable quark Yukawa coupling is given by 
\begin{equation}\label{Eq:Q1U3H1}
[b'a]+[a(\theta^{2}c)]+[(\theta^{2}c)b']
\qquad\rightarrow\qquad
W_{Q_{1}U_{3}H_{1}} \sim {\cal O}(1),
\end{equation}
and an exhaustive list of all leading quark Yukawa interactions can be found in table~\ref{YukawaForBaryons} of appendix \ref{App:B}.

Higher order interactions may appear in two ways. 
In the first case, non-renormalisable couplings in perturbation theory involve more than three D6-branes. One example is 
given by the sequence of four D6-branes,
\begin{equation}
[b'a]+[a(\theta^{2}a)]+[(\theta^{2}a)(\theta^{2}c)]+[(\theta^{2}c)b']
\qquad\rightarrow\qquad
W_{Q_{1}A_{2}U_{1}H_{1}} \sim {\cal O}(1),
\end{equation}
in which an adjoint representation $A_2$ of $SU(3)_a$ appears. Such non-renormalisable couplings are suppressed by the cut-off scale $M_{string}$
at high energies. Since the gravitational interaction of type IIA compactifications with D6-branes on toroidal orbifolds typically requires a high string scale~\cite{Blumenhagen:2003jy}, 
the non-renormalisable higher $n$-point interaction terms are negligible.

The second class of interaction terms, that is not taken into account in this article, is due to instantons on wrapped Euclidean D2-branes. 
Their strength is suppressed by $e^{-S_{D2}}$ and expected to be tiny in the perturbative regime of type II string theories~\cite{Blumenhagen:2009qh}.
Moreover, single instanton contributions only occur for a minimal number of fermionic zero modes. Since fractional three-cycles on $T^6/\Z_{2N}$
backgrounds are not rigid, all single D2-instanton contributions are absent in the model on $T^6/\Z_6'$.

\subsubsection{Breaking of the right-symmetric group $USp(2)_c \to U(1)_c$}\label{Sss:Breaking}

The three-point interaction terms discussed above and listed exhaustively at leading order in appendix~\ref{App:B} 
are invariant under the unbroken right-symmetric group
$USp(2)_{c}$. For example, in the previously derived interaction $Q_{1}U_{3}H_{1}$ in equation~(\ref{Eq:Q1U3H1}), the right-handed quarks and Higgses, 
\begin{equation}
U_{3}= \tarh{d_{3R}}{-u_{3R}}
\qquad\text{and}\qquad
H_{1}=\tarh{H_{1u}}{H_{1d}}
,
\end{equation}
form doublets under $USp(2)_{c}$.
Upon the breaking $USp(2)_{c}  \to U(1)_{c}$ by moving the D6-brane $(\theta c)$ away from the O6-plane  $(\sigma^2_c \neq 0)$ along $T^2_{(2)}$, 
the interaction terms (with contraction of the $USp(2)_c$ index by an epsilon tensor $U_3H_1\equiv \epsilon_{ij}U_3^iH_1^j$) split, e.g.
\begin{equation}
W_{Q_1U_3H_1} \; Q_{1}U_{3}H_{1} \quad \longrightarrow \quad W_{Q_1u_{3R}H_{1u}} \; Q_{1L}u_{3R}H_{1u} \; + \; W_{Q_1d_{3R}H_{1d}} \; Q_{1L}d_{3R}H_{1d}
,
\end{equation}
with $W_{Q_1u_{3R}H_{1u}} = W_{Q_1d_{3R}H_{1u}} \sim {\cal O}(e^{-\Delta\mathcal{A}_{(2)}(\sigma^2_c)})$ for $\sigma^2_a=\sigma^2_b=0$,
where $\Delta\mathcal{A}_{(2)}(\sigma^2_c)$ denotes the size of the triangle in dependence of the displacement $\sigma^2_c$.

\begin{figure}[!h]
\centering
\includegraphics[height=60mm]{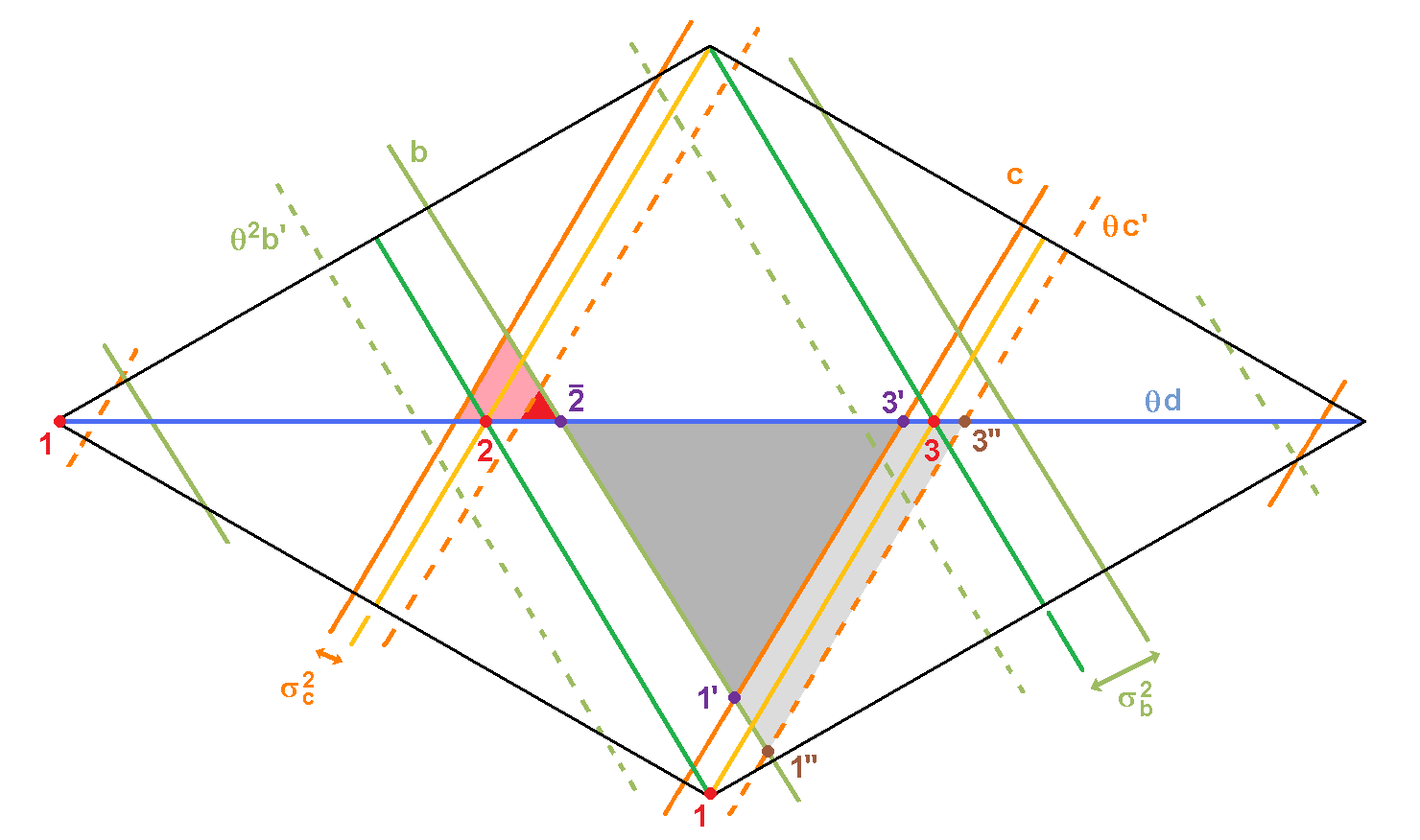}
\caption{\emph{Change of the size of triangular worldsheets upon continuous displacements $\sigma^2_b \neq \pm \sigma^2_c$ along the $\Z_2$ invariant
two-torus $T^2_{(2)}$. The middle lines in yellow and green correspond to the undisplaced D6-branes of figure~\protect\ref{LeptonsOnBranes},
while the solid and dashed lines correspond to the D6-branes $b$ and $(\theta^2 b')$ and $c$ and $(\theta c')$ in green and orange, respectively.}}
\label{DisplacementsofBranes}
\end{figure}

The vector-like matter states in the last two lines of equation~(\ref{FullSpectrum2}) are all rendered massive if also the stack of D6-branes $b$
is displaced along $T^2_{(2)}$ with $\sigma^2_b \neq \pm \sigma^2_c$. The situation is depicted in figure~\ref{DisplacementsofBranes}
for the lepton Yukawa intersections of D6-branes $b$, $c$ and $(\theta d)$ from section~\ref{Ss:Yukawa-SM},
leading to modifications of the sizes of all enclosed triangles with the D6-brane stacks $b$ or $c$ along some edge.
The change in size of the triangles can be computed e.g. for the case of the $[bc(\theta d)]$ sequence with displacements 
$0 \neq \sigma^2_b \neq \pm \sigma^2_c \neq 0, \sigma^2_d=0$, but no Wilson lines $\tau^2_x \equiv 0$ ($x=b,c,d$) from
\begin{equation}\label{Eq:Yukawas_complete_bcthetad}
\begin{aligned}
&\sum_{\mathcal{A}_{ijk,(2)}}e^{-\mathcal{A}_{ijk,(2)}/(2\pi\alpha')}  \sim \vartheta \targ{\delta_2}{0}(\frac{t_2\mathcal{A}_{(2)}}{2\pi}) = \sum_{l \in \Z} e^{-t_2\mathcal{A}_{(2)} (\delta_2 + l)^2} 
\quad
\text{with}
\quad 
t_2 = \frac{|I_{bc}^{(2)}I_{c(\theta d)}^{(2)}I_{(\theta d)b}^{(2)}|}{\gcd^2(I_{bc}^{(2)},I_{c(\theta d)}^{(2)},I_{(\theta d)b}^{(2)})} =3
, 
\\
\delta_2 &= \frac{i}{I_{bc}^{(2)}} + \frac{j}{I_{c(\theta d)}^{(2)}} + \frac{k}{I_{(\theta d)b}^{(2)}} 
+\frac{ \gcd(I_{bc}^{(2)},I_{c(\theta d)}^{(2)},I_{(\theta d)b}^{(2)}) \left( I_{bc}^{(2)} \sigma_d^2 + 
I_{c(\theta d)}^{(2)} \sigma_b^2 + I_{(\theta d)b}^{(2)} \sigma_c^2 \right) }{I_{bc}^{(2)} I_{c(\theta d)}^{(2)} I_{(\theta d)c}^{(2)}} + \frac{s(i,j,k)}{\gcd(I_{bc}^{(2)},I_{c(\theta d)}^{(2)},I_{(\theta d)b}^{(2)})}
\\
&=\frac{i+j+k}{3} + \frac{\sigma^2_b + \sigma^2_c}{3} + \frac{s(i,j,k)}{3}
,
\end{aligned}
\end{equation}
which corresponds to the Yukawa couplings in the non-coprime case~\cite{Cremades:2003qj} with \mbox{$I_{bc}^{(2)}=I_{c(\theta d)}^{(2)}=I_{(\theta d)b}^{(2)}=3$}.
In the absence of displacements, i.e. $\sigma^2_b = \sigma^2_c = \sigma^2_d =0$, the leading term  $i=j=k=0$  in the sum provides the 
unsuppressed Yukawa intersections for a given generation $E_nL_nH_n$ while for $i+j+k=3$  we recover the suppression factor of 1/6 of the two-torus area 
$\mathcal{A}_{(2)}$ of $T^2_{(2)}$ for the couplings like $E_1L_2H_3$ in figure~\ref{LeptonsOnBranes}.

With $\sigma^2_b \neq 0$ rendering all vector-like $\mathcal{N}=2$ supersymmetric matter on the last-but-one line 
as well as the antisymmetrics of $U(2)_b$  on the last line of equation~(\ref{FullSpectrum2}) massive and with $\sigma^2_c \neq 0, \pm \sigma^2_b$ 
leading to the breaking $USp(2)_c \to U(1)_c$ while rendering the remaining states on the last line of~(\ref{FullSpectrum2}) massive,
the couplings  $e_{n,R}L_nH_{n,d}$ to the right-handed charged leptons  become suppressed by a smaller area $\Delta \mathcal{A}_{(2)}(\sigma^2_b,\sigma^2_{c'})$ 
than those  $\nu_{n,R} L_nH_{n,u}$ to the right-handed neutrinos with $\Delta \mathcal{A}_{(2)}(\sigma^2_b,\sigma^2_c)$. 
The situation is depicted in figure~\ref{DisplacementsofBranes} as follows. 
The left-handed lepton $L_2$ is localised at the common apex of the two red and two grey triangles labelled $\ov{2}$. 
The right-handed electron $e_{2R}$ and down-type Higgs $H_{2d}$ sit at the remaining two apexes of the small dark red triangle, whereas the 
neutrino $\nu_{2R}$ and up-type Higgs $H_{2u}$ are supported at the two apexes of the bigger light triangle.

The area suppression of the non-diagonal couplings $e_{R1} L_2H_{3d}$ and $\nu_{R1} L_2H_{3u}$ changes in the opposite way, i.e. the former area is enlarged
while the latter is reduced, as illustrated in figure~\ref{DisplacementsofBranes} by the light and dark shaded areas with 
down-type Higgs $H_{3d}$ and $e_{3R}$ at the apexes $1''$ and $3''$ and  up-type Higgs $H_{3u}$ and $\nu_{3R}$ at the apexes $1'$ and $3'$.

Assuming that the {\it vev}s within one Higgs generation are of the same order, $\langle H_{1u}\rangle\approx\langle H_{1d}\rangle$,
 due to their origin from an $USp(2)_{c}$ doublet, the hierarchy of the masses of the up-type and down-type quarks  as well as of the right-handed charged leptons and neutrinos 
is generated solely by relative D6-brane displacements $\sigma^2_b, \sigma^2_c$ within a given particle generation.

%%%%%%%%%%%%%%%%%%%%%%%%%%%%%%%%%%%%%%%%%%%%%%%%%%%%%%%%%

\subsubsection{Family replication}\label{Sss:FamilyReplication}

In the lepton sector, three particle generations appear naturally because the diagonal couplings 
\begin{equation}
W_{E_{1}L_{1}H_{1}}, \quad W_{E_{2}L_{2}H_{2}}, \quad W_{E_{3}L_{3}H_{3}} \sim {\cal O}(1),
\end{equation}
in table~\ref{YukawaForLeptons} correspond to the only non-suppressed leptonic three-point interaction terms in the superpotential before 
switching on continuous displacements $\sigma^2_b,\sigma^2_c$. 
The non-holomorphic prefactor discussed in more detail in section~\ref{Ss:PhysicalYukawas} is universal for all these couplings, and
the mass hierarchy among the three lepton generations emerges from the choice of different {\it vev}s of the first three Higgs generations,
while the different Yukawa couplings within one lepton family are generated by the brane displacements $\sigma^2_b,\sigma^2_c$ discussed 
in section~\ref{Sss:Breaking}. 

There exist five types of subleading lepton Yukawa couplings with different volume suppression factors,
\begin{equation}
\begin{aligned}
W_{E_{i}L_{j}H_{k}}& \sim {\cal O}(e^{-\mathcal{A}_{(2)}/6})
\qquad
\text{with}
\qquad
(i,j,k) 
\quad
\text{permutations of }
(1,2,3)
,
\\
W_{E_{i}L_{i+3}H_{i+3}}& \sim {\cal O}(e^{-\mathcal{A}_{(1)}/12})
\qquad
\text{with}
\qquad
i=1,2,3
,
\\
W_{E_{i}L_{j}H_{k}}& \sim {\cal O}(e^{-\mathcal{A}_{(1)}/12-\mathcal{A}_{(2)}/6})
\qquad
\text{with}
\qquad
\left\{\begin{array}{rr}
i=1 & (j,k)=(5,6),(6,5) \\
2 & (6,4),(4,6)\\
3 & (4,5),(5,4)
\end{array}\right.
,
\\
W_{E_{i}L_{i+3}H_{i}}& \sim {\cal O}(e^{-\mathcal{A}_{(1)}/3})
\qquad
\text{with}
\qquad
i=1,2,3
,
\\
W_{E_{i}L_{j}H_{k-3}}& \sim {\cal O}(e^{-\mathcal{A}_{(1)}/3-\mathcal{A}_{(2)}/6})
\qquad
\text{with}
\qquad
\left\{\begin{array}{rr}
i=1 & (j,k)=(5,6),(6,5) \\
2 & (6,4),(4,6)\\
3 & (4,5),(5,4)
\end{array}\right.
,
\end{aligned}
\end{equation}
with the apexes of each triangle listed in table~\ref{YukawaForLeptons}.
The first kind leads to subleading mixings of the three generations in the lepton sector, which is flavour diagonal at leading order.
The second, third, fourth and fifth type of area suppressed couplings provide mixings with the additional vector-like
lepton pairs in equation~(\ref{FullSpectrum1}). These vector-like lepton pairs $(L_{3+i},\ov{L}_i)_{i=1,2,3}$ acquire masses through couplings
 to the symmetric representations $b_i$ of $U(2)_b$ as discussed below in section~\ref{Sss:Leptons}.

The situation is more complicated in the quark sector. As detailed in table~\ref{YukawaForBaryons} in appendix~\ref{App:B},
only two right-handed quark generations $U_1,U_3$ contribute to the non-suppressed holomorphic Yukawa couplings, which involve only one of the 
previously
considered Higgs generations, $H_1$, plus two other Higgs generations, where $H_4$ contributes to volume suppressed lepton Yukawa couplings,
and $H_7$ does not have any three-point interaction with leptons.  Moreover, the non-suppressed couplings involve the left-handed quark generations
$Q_1,Q_3,Q_4$ but not $Q_2$, where $Q_4$ and $Q_3$ provide the dominant contributions to the vector-like left-handed quark pair with admixtures of 
$Q_1$ and $Q_2$ as discussed below in section~\ref{Sss:Quarks}. The volume suppressed quark Yukawa couplings on the other hand involve three 
left-handed and two right-handed quark generations, $Q_2,Q_3,Q_4$ and $U_2,U_3$, as well as two of the `Standard Model' Higgs generations $H_2,H_3$ plus 
four further Higgs generations $H_5,H_6,H_8,H_9$. The three Higgs generations $H_7,H_8,H_9$ do not couple to the lepton sector, but only to the quark sector.

The situation in the quark sector becomes even more intricate when the three different kinds of non-holomorphic prefactors due to K\"ahler metrics
listed in the last column of table~\ref{YukawaForBaryons} are taken into account. A thorough analysis of the quark flavour structure therefore goes well 
beyond the scope of this article.

%%%%%%%%%%%%%%%%%%%%%%%%%%%%%%%%%%%%%%%%%%%%%%%%%%%%%%%%%%
\subsection{Physical Yukawa couplings and K\"ahler metrics}\label{Ss:PhysicalYukawas}

In section~\ref{Ss:Superpotential}, the holomorphic three-point couplings in the superpotential were discussed 
in terms of areas swept by worldsheet instantons. The physical Yukawa couplings $Y_{ijk}$ in the supergravity theory 
depend also on the K\"ahler potential ${\cal K}$ and on the open string matter K\"ahler metrics 
$K_{xy}$~\cite{Cremmer:1982en,Dixon:1989fj,Cvetic:2003ch,Abel:2003yx,Lust:2004cx},
\begin{equation}\label{Eq:Physical_Yukawas}
Y_{ijk} = \left( K_{xy} \, K_{yz} \, K_{zx} \right)^{-1/2} \; e^{\kappa_4^2 \; {\cal K}/2} \; W_{ijk}\; 
.
\end{equation}

While the factor $e^{\kappa_4^2 \; {\cal K}/2}$ containing the K\"ahler potential ${\cal K}$ is universal, the product of 
the K\"ahler metrics depends on each matter state and can potentially also contribute to relative suppressions 
of Yukawa couplings~\cite{Honecker:2011sm,Honecker:2011hm}. 
The individual formal expressions of the open string K\"ahler metrics have been computed at leading order 
in~\cite{Lust:2004cx,Akerblom:2007uc} and~\cite{Blumenhagen:2007ip} for three non-trivial intersection angles on the six-torus and 
$T^6/\Z_2\times \Z_2$ without and with discrete torsion, respectively, and the missing cases with one or three vanishing angles on 
arbitrary orbifolds were added in~\cite{Honecker:2011sm,Honecker:2011hm}, where it was also explicitly shown that the K\"ahler metrics 
do not depend on Wilson line or displacement moduli and have identical expressions - up to the one exception of a different normalisation 
for adjoint matter on parallel D6-branes - for all types of bulk, fractional and rigid D6-branes on the six-torus, $T^6/\Z_{2N}$ and 
$T^6/\Z_2 \times \Z_{2M}$ orbifold backgrounds with discrete torsion, respectively. 
Recently, it was shown in~\cite{Berg:2011ij} for the six-torus and $T^6/\Z_6'$ orbifold
(based on computations of two-point correlators of chiral fields on the six-torus in~\cite{Benakli:2008ub})
that the one-loop corrections to the K\"ahler metrics of adjoints from ${\cal N}=1$ supersymmetric sectors vanish,
while those from ${\cal N}=2$ supersymmetric sectors modify the definition of the K\"ahler moduli.
All formal expressions (at leading order) for open string K\"ahler metrics are displayed in table~\ref{tab:Kaehler_Metrics_General}.
\begin{table}[ht!]
\begin{center}
{\small 
\begin{equation*}
\begin{array}{|c||c|c|c|}\hline
\multicolumn{4}{|c|}{\text{\bf K\"ahler metrics and intersection angles}}
\\\hline\hline
(\vec{\phi}_{ab}) & (0,0,0) & (0^{(i)},\phi^{(j)},\phi^{(k)}) & (\phi^{(1)},\phi^{(2)},\phi^{(3)})
\\\hline
K_{\mathbf{R}_a} & \begin{array}{lr}
\frac{\sqrt{2 \pi}}{c_a} \, \frac{f(S,U)}{v_i} \sqrt{\frac{V_{aa}^{(j)} V_{aa}^{(k)}}{V_{aa}^{(i)}}} & {\Adj_a}
\\  f(S,U) \;   \sqrt{\frac{ 2 \pi V_{ab}^{(i)}}{v_jv_k}} 
&{\mathbf{R}_a \neq \Adj_a} 
\end{array} & 
f(S,U) \;   \sqrt{\frac{ 2 \pi V_{ab}^{(i)}}{v_jv_k}}
& \frac{f(S,U)}{\sqrt{v_1v_2v_3}}  \,  \sqrt{\prod_{i=1}^3 
\left(\frac{\Gamma(|\phi^{(i)}_{ab}|)}{\Gamma(1-|\phi^{(i)}_{ab}|)}\right)^{-\frac{\sgn(\phi^{(i)}_{ab})}{\sgn(I_{ab})} }}
\\\hline
\end{array}
\end{equation*}
}
\end{center}
\caption{K\"ahler metrics for open strings at arbitrary supersymmetric intersection angles. 
The K\"ahler modulus $v_i$ denotes the two-torus volume of $T^2_{(i)}$.
$\sqrt{V_{ab}^{(i)}}$ parameterises the one-cycle volume of D6-branes parallel along $T^2_{(i)}$ normalised by $1/\sqrt{v_i}$ such that it only
contains a complex structure moduli dependence, 
and the factor $f(S,U)=(S\, \prod_{i=1}^{h_{21}^{\rm bulk}} U_i)^{-\alpha/4}$ contains the universal dependence on the 
four-dimensional field theoretical dilaton $S$ and bulk complex structure moduli $U_i$ where $h_{21}^{\rm bulk}=1$ and $\alpha=2$
for the $T^6/\Z_6'$ background of this article. The K\"ahler metrics for adjoint matter carry the normalisation
factor $c_a=1,2$ on stacks of identical  bulk and fractional D6-branes, respectively, whereas all other K\"ahler metrics
are independent of the type of (bulk, fractional or rigid) D6-brane.}
\label{tab:Kaehler_Metrics_General}
\end{table}
The prefactor $\left( K_{xy} \, K_{yz} \, K_{zx} \right)^{-1/2}$ in equation~(\ref{Eq:Physical_Yukawas}) to the physical Yukawa coupling depends thus on 
the three relative intersection angles $(\vec{\phi}_{xy})$, or some one-cycle length $\sqrt{V^{(i)}_{xy}}$ and K\"ahler moduli $v_i$ for some vanishing angle $\phi_{xy}^{(i)}=0$, 
all of which can in principle contribute as well to a hierarchical structure of  the Yukawa couplings. In tables~\ref{YukawaForLeptons} 
to~\ref{Hidden-Sequences} in appendix~\ref{App:B} both suppression factors due to the size of closed triangles
and different K\"ahler metrics are displayed in the second and last column of each table, respectively, for all possible (charged) three-point interactions. 
As an example, all leptonic Yukawa couplings in table~\ref{YukawaForLeptons} carry the same prefactor from the K\"ahler metrics
with $\frac{1}{2(50)^{1/4}} \approx 0.188$, but differ in the size of enclosed triangles, whereas the quark Yukawa couplings in table~\ref{YukawaForBaryons}
 have both different factors from K\"ahler metrics with $\frac{1}{10^{3/4}} \approx 0.178$ and $\frac{1}{5 \cdot 2^{1/4}} \approx 0.168$ and 
$\frac{3^{1/8}}{2(25\pi v_1)^{1/4}}\approx \frac{0.193}{v_1^{1/4}}$ and different sizes of triangular worldsheets. While the numerical factors differ, the 
order of magnitude is identical for these examples, except for the volume dependence $v_1^{-1/4}$ in the last case, which might lead to a significant 
suppression of the corresponding couplings.
To firmly establish such an extra volume suppression of physical Yukawa couplings, it will, however, be necessary to 
derive the holomorphic worldsheet instanton contributions in equation~(\ref{ThreePointInteraction}) for D6-branes at one vanishing angle on $T^6/\Z_{2N}$
from first principles and thereby exclude the possibility of cancellations in the $v_i$ dependence among the two factors in the physical Yukawa 
interactions~(\ref{Eq:Physical_Yukawas}).

One might speculate that the same order of magnitude of the prefactors (except for the $v_1^{-1/4}$ dependence)
is due to the limited choice of supersymmetric three-cycles on $T^6/\Z_6'$, which do 
not exceed the RR tadpole cancellation conditions, and correspondingly a small set of relative intersection angles and normalised one-cycle volumes.

%%%%%%%%%%%%%%%%%%%%%%%%%%%%%%%%%%%%%%%%%%%%%%%%%%%%%%%%%%%%%%%%%%%%%%%%%%%%%%%%%%%%

\section{Masses of non-chiral representations}\label{Ss:masses}

According to the method described in the previous section, we can proceed to list all leading order three-point couplings of the Standard Model 
with hidden $USp(6)_h$ on $T^6/\Z_6'$. The complete list can be found in tables~\ref{YukawaForLeptons} to~\ref{Hidden-Sequences} in appendix~\ref{App:B}. 
While the Yukawa couplings in the lepton and quark sector have been discussed in detail in section~\ref{Ss:Yukawa-SM}, in this section the 
couplings of the vector-like matter representations and mass terms through a Higgs-like mechanism involving {\it vev}s 
inside the (singlets in the decomposition into irreducible representations under $SU(N_x) \times U(1)_x$ of the)
adjoint matter representations of $U(3)_a$ and $U(2)_b$ as well as inside the symmetrics of $U(2)_b$ are discussed.

\subsection{The vector-like fields in the particle spectrum}\label{The abundant fields in the particle spectrum}

The massless matter spectrum was given in section \ref{The full spectrum} with localisations at D6-brane intersections listed in appendix~\ref{App:A}
 in tables~\ref{IntersectionNumbers-1} and~\ref{IntersectionNumbers-2} for adjoint and bifundamental and symmetric and antisymmetric representations, respectively. 
Besides the three Standard Model quark and lepton generations, the model contains several representations that are vector-like with respect to the Standard Model
gauge group. These can be classified as follows:
\begin{enumerate}
\item
The `chiral' spectrum in equation~(\ref{FullSpectrum1}) contains three vector-like lepton pairs at $b(\theta d)$ plus $b(\theta^2 d')$ 
intersections as well as six plus three Higgs generations at $bc$ and $b(\theta c)$ intersections, respectively, which are counted as `chiral' with respect to the 
massive anomalous $U(1)_b \subset U(2)_b$ symmetry. Leptons and Higgses are clearly distinguished by the anomaly-free $Q_{B-L} = \frac{1}{3} Q_a + Q_d$ charge.
Possible mass terms for the vector-like leptons and Higgses via three-point couplings to matter in the symmetric representation of $U(2)_b$
are discussed below in section~\ref{Sss:Leptons}.
\item
The last two lines of the `non-chiral' spectrum in equation~(\ref{FullSpectrum2}) consist of $\mathcal{N}=2$ supersymmetric
hypermultiplets at D6-branes $x$ and $y$ parallel along the $\Z_2$ invariant $T^2_{(2)}$, and continuous
relative D6-brane displacements render the full $\mathcal{N}=2$ multiplets massive, cf. equation~(\ref{Eq:M-parallel}),
 which is denoted by the lower index $m$ of the multiplicities in equation~(\ref{FullSpectrum2}) and labelled by $|\cdot|_{m}$ in tables~\ref{IntersectionNumbers-1} 
and~\ref{IntersectionNumbers-2}. States with such a manifest supersymmetric mass term will not be further discussed in this section.
\item
The first two lines of the `non-chiral' spectrum in equation~(\ref{FullSpectrum2}) and the `hidden' spectrum in equation~(\ref{FullSpectrum3})
comprise:
(i) one left-handed vector-like quark pair with two different possible mass terms at leading order discussed below in section~\ref{Sss:Quarks},
(ii) three kinds of vector-like right-handed quark pairs with exotic $B-L$ charge, for which mass terms are discussed in section~\ref{Sss:VectorLike},
(iii) adjoint matter of $U(3)_a$ and $U(2)_b$ as well as vector-like pairs of symmetric matter of $U(2)_b$, for which {\it vev}s are discussed in
 section~\ref{Ss:Adjoint-vevs} and~\ref{Vacuum expectation value of (anti)symmetric representations}, respectively,
(iv) matter charged under the hidden gauge group $USp(6)_c$ as briefly discussed in section~\ref{S:Couplings_hidden}.
\end{enumerate}

In the following, mass terms via three-point couplings of the vector-like representations in items 1 and 3, 
which originate from two distinct microscopically ${\cal N}=1$ supersymmetric sectors in conjugate representations
with respect to the Standard Model gauge group, are discussed with a Standard Model singlet field, which obtains a {\it vev}.

%%%%%%%%%%%%%%%%%%%%%%%%%%%%%%%%%%%%%%%%%%%%%%%%%
\subsubsection{Vector-like lepton and Higgs pairs}\label{Sss:Leptons}

The `chiral' spectrum in equation~(\ref{FullSpectrum1}) contains six leptons  $L_i=(\1,\2)_{-1/2}$
and three anti-leptons $\ov{L}_i=(\1,\2)_{1/2}$, which according to table~\ref{IntersectionNumbers-1} are localised at $b(\theta d)$ and $b(\theta^2 d')$ 
intersections. Due to the same anomalous $U(1)_b \subset U(2)_b$ charge, the three vector-like combinations do not have perturbative three-point couplings 
to adjoints $B_i$ of $U(2)_b$, but instead charge selection rules  allow for perturbative couplings to the conjugate 
symmetric representations $b_i=(\1,\ov{\3}_{\Sym})_0$ of $U(2)_b$ in the second line of equation~(\ref{FullSpectrum2}).
Similarly, the anomalous $U(1)_b$ charge of the Higgs multiplets $H_d^i=(\1,\ov{\2})_{-1/2}$ and $H_u^i=(\1,\ov{\2})_{1/2}$
only admits three-point interactions with the symmetrics $\ov{b}_i = (\1,\3_{\Sym})_0$ of $U(2)_b$ at the perturbative level.
The localisation of all symmetric and antisymmetric matter states of $U(2)_b$ are displayed in table~\ref{tab:local_Sym_U2}.
%%%%%%%%%%%%%%%%%%%%%%%%%%%%%%
\begin{table}[ht!]
\begin{center}
\setlength{\tabcolsep}{10pt}
\begin{tabular}{|c||c|c|c|}
\hline
\muc{4}{|c|}{\text{\bf Localisation of symmetric and antisymmetric matter of $U(2)_b$}}
\\\hline\hline
$b(\theta^{k}b')$ & representation
& \begin{tabular}{c}intersection point\\$\{T_{(1)},T_{(2)},T_{(3)}\}$\\\end{tabular} & field label
\\ \hline\hline
$bb'$          & $(\1,\ov{\3}_{\Sym})_0$ & $\{(R,R'),i,3\}$     & $b_{i}$ with $i=1,2,3$   \\
                       &       & $\{(R,R'),i,1\}$     & $b_{3+i}$ with $i=1,2,3$   \\\hline
$b(\theta b')$    & $(\1,\3_{\Sym})_0$ & $\{\parallel=,i,3\}$ & $\bar{b}_{i}$  with $i=1,2,3$  \\
                      &       & $\{\parallel=,i,1\}$ & $\bar{b}_{3+i}$  with $i=1,2,3$   \\ \hline
$b(\theta^{2}b')$  & $[(\1,\1_{\Anti})_0+c.c.]_m  $ &  $\begin{array}{c} \{5,\parallel \neq ,k\}, \{(S,S'),\parallel \neq ,k\} \\ \text{with } k=1,3 \end{array}$  &                        \\ \hline
\end{tabular}
\end{center}
\caption{Localisation of symmetric and antisymmetric matter at intersection points of the D6-brane $b$ and the orientifold images $(\theta^k b')$.
Only labels for the fields with three-point couplings to vector-like lepton or Higgs pairs are given.}
\label{tab:local_Sym_U2}
\end{table}
%%%%%%%%%%%%%%%%%%%

Table~\ref{LLbarSym} in appendix~\ref{App:B} shows that the first three lepton generations $L_1,L_2,L_3$ do not contribute to the couplings
without area suppression, confirming their interpretation in section~\ref{Sss:FamilyReplication} as dominant contributions to the three light physical 
lepton generations. The couplings to the conjugate symmetric representations $b_i$ fall into two classes,
\begin{equation}\label{Eq:L+H_Suppressions}
\begin{aligned}
W_{\ov{L}_i L_{3+i}, b_i} & \sim {\cal O}(1) 
,
\\
W_{\ov{L}_i L_j b_k} & \sim {\cal O}(e^{-\sum_{k=1}^{3}\mu_{k}\mathcal{A}_{(k)}}) 
\quad \text{ with }\quad \vec{\mu}=\left(0\text{ or }1/6;0\text{ or }1/6;0\text{ or }1/4\right)\neq\vec{0}
,
\end{aligned}
\end{equation}
as detailed in the upper half of table~\ref{LLbarSym},
which means that at leading order, massive vector-like leptons are given by the diagonal pairs $(\ov{L}_i,L_{3+i})_{i=1,2,3}$ with tiny mixings 
among each other and with the $(L_{i})_{i=1,2,3}$ due to the volume suppressed couplings.

The situation is similar for the six Higgs generations $H_i, H_{3+i}$ at $bc$ intersections with $i\in \{1,2,3\}$ labelling their localisation along $T^2_{(2)}$.
Their three point interactions with the symmetrics $\ov{b}_i$ of $U(2)_b$ are listed at leading order in the lower half of table~\ref{LLbarSym}, 
displaying similar suppression factors as the leptons in equation~(\ref{Eq:L+H_Suppressions}). The interpretation is slightly different due to the absence
of `diagonal' mass terms. However, we expect mass eigenstates to be formed by linear combinations of the six Higgs generations $H_i,H_{3+i}$. 

For the remaining three Higgs generations $H_{6+i}$ at $b(\theta c)$ intersections, there exists no sequence involving three D6-branes.
  Nevertheless, couplings of the form $H_{i}a_{j}H_{k}$, with $i=7,8,9$; $k=1,...,6$ and $a_{j}$ one of the (massive) antisymmetric representations of $U(2)_b$
are allowed by charge selection rules and closed sequences. Also couplings like $H_{i}\bar{b}_{j}\tilde{H}$ with $i=7,8,9$ and $\tilde{H}$ being one of the (massive) 
non-chiral representations like the Higgses at the intersections $b(\theta^{2}c)$ may exist. Such a term contributes terms of the form
$(H\bar{b})^2 + (\bar{b}H)^2 + (H \tilde{H})^2$ to the scalar potential, and therefore a {\it vev} of $\bar{b}_i$ could still generate 
  masses for the Higgses. 
Since the three Higgs generations $H_{6+i}$ do not have lepton Yukawa couplings and only suppressed quark Yukawas, we do not further investigate 
their masses at this point.

In section~\ref{Vacuum expectation value of (anti)symmetric representations} the conditions for {\it vev}s in the symmetric representations and their 
effects on gauge symmetry breaking of $SU(2)_b \subset U(2)_b$ are discussed further.

%%%%%%%%%%%%%%%%%%%%%%%%%%%%%%%%%%%%%%%%%%%%%%%%%5
\subsubsection{Left-handed vector-like quark pair}\label{Sss:Quarks}

The massless spectrum contains four left-handed quarks $Q_{i}$  at $ab'$ and $a(\theta b')$ intersections 
and one left-handed anti-quark $\ov{Q}$ at an $a(\theta^2 b')$ intersection (cf. table~\ref{IntersectionNumbers-1}),
for which  a gauge invariant mass term $W=\sum_i m_i Q_i \ov{Q}$ exists in field theory. But such a term is 
in type IIA string compactifications only expected for microscopic ${\cal N}=2$ supersymmetric sectors.
There exist, however, several three point interaction terms between the left-handed quarks $Q_i$, $\ov{Q}$ and adjoint representations  $A_j$ of $U(3)_a$ and 
$B_k$ of $U(2)_b$ listed in table~\ref{QQbarU2} of appendix~\ref{App:B}. 
The localisations of adjoints $B_k$ of $U(2)_b$ are given in table~\ref{tab:local_Adj_U2}, while $A_1$ off $U(3)_a$ stems from the $aa$ sector, and 
$A_2$ is located in the only intersection of orbifold images $a(\theta^k a)_{k \in \{1,2\}}$ along $T^2_{(1)} \times T^2_{(2)}$.
%%%%%%%%%%%%%%%%%%%
\begin{table}[ht!]
\begin{center}
\setlength{\tabcolsep}{10pt}
\begin{tabular}{|c||c|c|c|}
\hline
\muc{4}{|c|}{\text{\bf Localisation of adjoint matter of $U(2)_b$}}
\\\hline\hline$b'(\theta^{k}b')$  & representation & \begin{tabular}{c}intersection point\\$\{T_{(1)},T_{(2)},T_{(3)}\}$\\\end{tabular} & field  label
\\ \hline\hline
$b'b'$           &  $(\1,\3_{\Adj})_0+(\1,\1)_0$ & not localised         & $B_{1}$   \\ \hline
$b'(\theta^k b')_{k=1,2}$    &  $(\1,\3_{\Adj})_0+(\1,\1)_0$ & $\{4,i,\parallel=\}$  & $B_{1+i}$ with $i=1,2,3$   \\
                    &      & $\{(R,R'),i,\parallel=\}$  & $B_{4+i}, B_{7+i}$ with $i=1,2,3$  \\ \hline
\end{tabular}
\end{center}
\caption{Localisation of adjoint matter of $U(2)_b$ at intersection points of the D6-branes $b'$ and its orbifold images $(\theta^k b')_{k=1,2}$.
The parameterisation is chosen to fit the discussion of masses for symmetric matter in section~\protect\ref{Sss:VectorSymmetrics}.
}
\label{tab:local_Adj_U2}
\end{table}
%%%%%%%%%%%%%%%%%%%%%%%%
Two couplings, $\ov{Q}Q_{4}A_{2}$ and  $\ov{Q}Q_{3}B_{2}$, are not volume suppressed, which means that if the singlets (cf.
the decomposition in equation~(\ref{Eq:Decompose_Adjoints_UN}))
inside the adjoint representations $A_2$ and $B_2$ at intersections of orbifold image D6-branes 
$a(\theta^k a)_{k \in \{1,2\}}$ and $b(\theta^k b)_{k \in \{1,2\}}$ receive {\it vev}s, a linear combination of $Q_4$ and $Q_3$ will form a massive quark pair with $\ov{Q}$,
and only three Standard Model chiral quark generations remain massless. This statement remains true even if the volume suppressed three-point couplings 
in table~\ref{QQbarU2} are taken into account, which lead to a mixing of the dominant contributions to the massive left-handed quark,  $Q_4$ and $Q_3$, 
 with tiny admixtures of $Q_1$ and $Q_2$.
Yukawa couplings of the three massless Standard Model quark generations have already been discussed in section~\ref{Sss:FamilyReplication}.

Details on flat directions of {\it vev}s of the singlets inside the adjoint representations of $U(3)_a$ and $U(2)_b$ are discussed below in section~\ref{Ss:Adjoint-vevs}.

%%%%%%%%%%%%%%%%%%%%%%%%%%%%%%%%%%%%%%%%%%%%%%%%%5
\subsubsection{Right-handed vector-like quark exotics}\label{Sss:VectorLike}

The second line of the `non-chiral' spectrum~(\ref{FullSpectrum2}) contains three kinds of vector-like right-handed quark pairs with
exotic $B-L$ charge labelled $Y_1,Y_3,Y_5$ for $(\3,\1)_{-1/3}$ and $Y_2,Y_4,Y_6$ its conjugate representation at $a(\theta^k d)_{k \in \{0,1,2\}}$
intersections, $Z_1,Z_3,Z_5$ for $(\3,\1)_{2/3}$ and $Z_2,Z_4,Z_6$ its conjugate representation at $a(\theta^k d')_{k \in \{0,1,2\}}$
intersections, $W_1,W_3,W_5$ for $(\ov{\3}_{\Anti},\1)_{1/3}$ and $W_2, W_4,W_6$ for its conjugate representation at $a(\theta^k a')_{k \in \{0,1,2\}}$ intersections.
Details on the localisation in a given intersection sector can be found in table~\ref{IntersectionNumbers-1} for the bifundamental representations $Y_i$, $Z_i$
and in table~\ref{IntersectionNumbers-2} for the antisymmetric representations $W_i$.

All bifundamental representations $Y_i$, $Z_i$ and antisymmetrics $W_i$ couple to the adjoint representation
$A_2$ of $U(3)_a$ at the $a(\theta^k a)_{k \in \{1,2\}}$ intersection with various volume suppression factors along $T^2_{(1)} \times T^2_{(2)}$ 
as listed in table~\ref{tab:qR_to_adjoints}. 
In analogy to the example of~\cite{Cremades:2003qj}, the three-point interactions involving the adjoint $A_2$ can be expressed, e.g.  as $\sum_{i,j=1}^3 W_{Y_{2i-1} Y_{2j} A_2} \, Y_{2i-1} Y_{2j} A_2$, 
with the coefficients $\left(W_Y\right)_{ij} \equiv W_{Y_{2i-1} Y_{2j} A_2} $ in the factorised form
\begin{equation}
W = A \cdot X \cdot B 
\end{equation}
with $A$ and $B$ diagonal $3 \times 3$ matrices and $X$ having entries 0 or 1 only, in detail
\begin{equation}
\begin{aligned}
X_Y =&   \left(\begin{array}{ccc} 0 & 0 & 1 \\ 0 & 0 & 1 \\ 1 & 1 & 1 \end{array}\right)
\qquad \text{ and } \qquad
A_Y=B_Y=  {\rm diag} \left(e^{-{\cal A}_{(1)}/16}  , e^{-{\cal A}_{(1)}/16-{\cal A}_{(2)}/4} , 1 \right)
,
\\
X_{Z} =& \left(\begin{array}{ccc} 1 & 1 & 1 \\ 0 & 0 & 1 \\ 0 & 0 & 1 \end{array}\right)
\qquad \text{ and } \qquad
\left\{\begin{array}{c}
A_Z= {\rm diag} \left(1, e^{-{\cal A}_{(1)}/16} , e^{-{\cal A}_{(1)}/16-{\cal A}_{(2)}/4}  \right)\\
B_Z= {\rm diag} \left(e^{-{\cal A}_{(1)}/16}, e^{-{\cal A}_{(1)}/16-{\cal A}_{(2)}/4} ,1 \right)
\end{array}\right.
,
\\
X_{W} =&  \left(\begin{array}{ccc} 1 & 1 & 1 \\ 1 & 0 & 0 \\ 1 & 0 & 0 \end{array}\right)
\qquad \text{ and } \qquad
A_W=B_W = {\rm diag} \left(e^{-{\cal A}_{(1)}/16} ,   e^{{\cal A}_{(1)}/16} ,   e^{{\cal A}_{(1)}/16} \right)
.
\end{aligned}
\end{equation}
Upon $A_2$ acquiring a {\it vev}, two out of three right-handed vector-like quark pairs of each kind $Y_i,Z_i,W_i$
are thus rendered massive simultaneously with the left-handed vector-like quark pair $\ov{Q}Q_4$ discussed in section~\ref{Sss:Quarks}.
The third generation of each kind acquires a mass as follows: two
vector-like combinations of antisymmetric representations $(W_j,W_k)_{j \in\{3,5\},k \in \{4,6\}}$ are localised in the  $a(\theta a')$ sector of parallel orientifold image D6-branes. 
In contrast to various bifundamental sectors as well as the symmetrics in the $c(\theta c')$ sector,
a displacement $\sigma^2_a \neq 0$ from the origin does not provide a mass term, since the stack of D6-branes $a$ is perpendicular to the orbit of O6-planes 
along $T^2_{(2)}$, and $a'$ will have the same displacement, $\sigma^2_{a'} =\sigma^2_a$, while $\sigma^2_{x'}=-\sigma^2_x$ for $x \in \{b,c,d\}$.
This is in agreement with the fact that $USp(6)_h$ with $h$ parallel to $a$ cannot be broken by a displacement $\sigma^2_h$~\cite{Gmeiner:2008xq}.
To keep the number of displacements minimal while rendering ${\cal N}=2$ supersymmetric sectors massive, we set $\sigma^2_a=0$ throughout this article.
At this point, we make the ansatz of coupling $(W_j,W_k)_{j \in\{3,5\},k \in \{4,6\}}$ to the adjoint representation $A_1$ of the $aa$ sector.
This results in the three-point interactions without volume suppression on the first line of table~\ref{tab:qR_to_adjoints}.
Similarly, the vector-like combinations $(Y_1,Y_2)$ and $(Y_3,Y_4)$ arise each from a pair (under $\Z_2$) of  $ad$ intersection points along $T^2_{(1)}$.
The same is true for $(Z_2,Z_3)$ and $(Z_4,Z_5)$ at $a(\theta d')$ intersections. For each of these pairs, couplings to $A_1$ are  allowed by the two selection rules
of charge neutralness and closed triangles. These three-point couplings without volume suppression are included on the last two rows of $Y_i$ and on the last two lines
of $Z_i$ couplings in table~\ref{tab:qR_to_adjoints}.
Alternatively, the same vector-like pairs $(Y_{2i-1},Y_{2j})$ and $(Z_{2i-1},Z_{2j})$ couple to the singlet field in the `adjoint' $D_1$ of $U(1)_d$ from the $dd$ sector
without area suppression and to the $D_{2 \ldots 10}$ at $d(\theta^k d)_{k \in \{1,2\}}$ intersections with various suppression factors along $T^2_{(1)} \times T^2_{(2)}$. 
A {\it vev} of any $D_k$ does, however, not provide the missing mass term for the third vector-like right-handed $W_i$ quark generation.

All vector-like quarks on the second line of~(\ref{FullSpectrum2}) are thus rendered massive by 
three-point couplings to two adjoints $A_1$ and $A_2$ from the $aa$ and $a(\theta^k a)_{k\in \{1,2\}}$ sectors, respectively, if some {\it vev}s 
for both states $A_1, A_2$ are generated.

%%%%%%%%%%%%%%%%%%%%%%%%%%%%%%%%%%%%%%%%%%%%%%%%%5
\subsubsection{Vector-like symmetrics of $U(2)_b$}\label{Sss:VectorSymmetrics}

The vector-like symmetric representations of $U(2)_b$ are composed of states $b_i=(\1,\ov{\3}_{\Sym})_0$ at $bb'$
and $\ov{b}_i=(\1,\3_{\Sym})_0$ at $b(\theta b')$ intersections as detailed in table~\ref{tab:local_Sym_U2}.
By the selection rule of closed triangular worldsheets, they can acquire masses via couplings to 
the adjoints $B_i$ at $b(\theta^k b)_{k\in \{1,2\}}$ intersections listed in table~\ref{tab:local_Adj_U2}.
The complete list of leading order couplings is given in table~\ref{tab:Sym+Adj_U2b} of appendix~\ref{App:B}
with the following result. There exist two distinct sets $(b_i,\ov{b}_j)$ with $i,j \in \{1,2,3\}$
and $i,j\in \{4,5,6\}$ among which no couplings exist due to the lack of triangles on $T^2_{(3)}$. 
At leading order with no area suppression, the three-point couplings are diagonal, $i=j$, with both adjoints
$B_{4+i}$ and $B_{7+i}$ of $U(2)_b$ located at the same intersection points. Further area suppressed 
couplings lead to mixings within the two distinct sets of symmetric representations.
While both couplings to the adjoints $B_{4+k}$ and $B_{7+k}$ appear at our level of discussion on equal footing,
it might be possible that a fully string theoretic computation leads to further selection rules e.g. taking
into account the different $\Z_2$ transformations of the two massless states per intersection.
Such a selection rule does not exist for the six-torus, for which the holomorphic Yukawa interactions
were derived~\cite{Cremades:2003qj,Cremades:2004wa}, but a detailed investigation goes beyond the scope of this article.

This completes the discussion of possible mass terms of charged vector-like states in the bifundamental, antisymmetric and symmetric representation 
 for the Standard Model with `hidden' $USp(6)_h$ on $T^6/\Z_6'$. It remains to show on the one hand in sections~\ref{Ss:Adjoint-vevs} 
and~\ref{Vacuum expectation value of (anti)symmetric representations} that the assumed {\it vev}s of adjoint and symmetric matter  representations 
of $U(3)_a$ and $U(2)_b$ preserve supersymmetry and gauge symmetry,  and on the other hand that these {\it vev}s also provide masses for the adjoint 
matter states of $SU(3)_a$ and $SU(2)_b$ on the first line of~(\ref{FullSpectrum2}) themselves.  

Finally, in section~\ref{S:Couplings_hidden} we briefly comment on the states with hidden sector charges.

%%%%%%%%%%%%%%%%%%%%%%%
\subsection{Vacuum expectation values of adjoint representations of $U(3)_a$ and $U(2)_b$}\label{Ss:Adjoint-vevs}

In section~\ref{The abundant fields in the particle spectrum}, it was argued that all vector-like quark states acquire mass terms by couplings to 
matter in the adjoint representation of $U(3)_a$, or for the left-handed vector-like quark pair also to the adjoint representation of $U(2)_b$.

Supersymmetry requires the {\it vev}s inside these adjoint representations to satisfy both the $D$- and $F$-term constraints, and 
in order to avoid a breaking of the gauge groups, the {\it vev}s have to be imposed on the singlet contribution,
\begin{equation}\label{Eq:Decompose_Adjoints_UN}
({\bf 9}_{U(3)_a}) = ({\bf 8}_{SU(3)_a})_0 + (\1)_0,
\qquad
({\bf 4}_{U(2)_b}) = (\3_{SU(2)_b})_0 + (\1)_0,
\end{equation}
upon the decomposition $U(N_x) = SU(N_x) \times U(1)_x$.

%%%%%%%%%%%%%%%%%%%%%%%%%%%%%%%%%%%%%%%%
\subsubsection{Vacuum expectation values in the adjoints of $U(3)_a$}\label{Sss:Adjoints_U3a}

Due to the reality of the adjoint representation $A$, the $D$-term contribution from such a field vanishes identically,
\begin{equation}
D^{a}(A) =-g\left(\text{Tr}(A^{\dagger}t^{a}A)-\text{Tr}(At^{a}A^{\dagger})\right) =0
,
\end{equation}
where $t_i=t_i^{\dagger}$ are the generators of $SU(N_a) \times U(1)_a$ with $t_0=\unity/\sqrt{N_{a}}$ and $A = \sum_{i=0}^{N_a^2-1} A^i t_i$ for the adjoint 
representation.

\begin{figure}[!h]
\centering
\includegraphics[width=12cm]{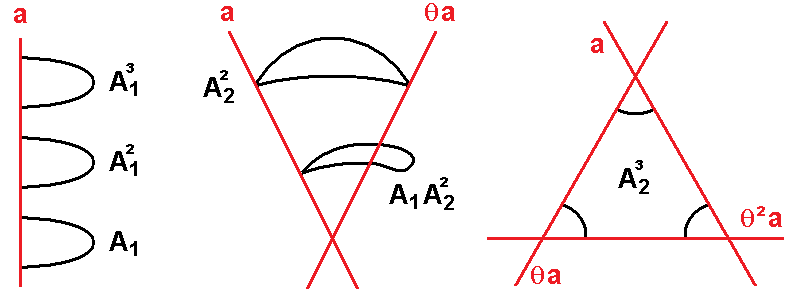}
\caption{\emph{Intersections of the D6-branes $a$, with its orbifold images $(\theta a)$ and $(\theta^{2} a)$ and the couplings for the adjoint representations $A_1$ and $A_2$
of $U(3)_a$ which are allowed by the existence of closed polygons.}}
\label{SuperpotentialAdjoints}
\end{figure}

While the $D$-term condition is trivially fulfilled, the $F$-term condition relates the {\it vev}s of the two singlet fields pertaining to respectively the 
adjoint multiplets $A_1$ and $A_2$ on parallel D6-branes, $aa$, and at the intersection of orbifold images, $a(\theta^k a)_{k \in \{1,2\}}$, 
of the Standard Model with hidden $USp(6)_h$ on $T^6/\Z_6'$ as follows. Analogously to the string selection rules of closed polygons used in 
sections~\ref{Ss:Yukawa-SM} and~\ref{The abundant fields in the particle spectrum} to determine all three-point interaction terms with bifundamental, 
symmetric or antisymmetric matter, we make the ansatz for the general form of the superpotential terms involving only the adjoint fields $A_1$, $A_2$
as depicted in figure~\ref{SuperpotentialAdjoints},
\begin{equation}
W=\xi \, \text{Tr}(A_{1})+\gamma \, \text{Tr}(A_{1}^{2})+\frac{\mu}{3} \, \text{Tr}(A_{1}^{3})+\lambda \, \text{Tr}(A_{2}^{2})+\frac{\beta}{3} \, \text{Tr}(A_{2}^{3})+\alpha \, \text{Tr}(A_{1}A_{2}^{2})
,
\label{SuperpotentialAdjoint}
\end{equation}
with unknown constants $\xi, \gamma, \mu, \lambda, \beta, \alpha$.
The first term only receives contributions from the singlet inside the adjoint of $U(3)_a$.\footnote{
Strictly speaking, each coefficient depends on (untwisted) closed string moduli, just like the three-point interactions discussed before. 
Throughout the section we assume that the relevant moduli take finite constant values. \\
In~\cite{Lust:2004cx,Blumenhagen:2006ci} based on earlier work of the heterotic string~\cite{Dixon:1989fj}, it was argued that $\xi=0$ in perturbation theory
due to the existence of a conserved Abelian worldsheet current  $U(1)_{WS}$, which serves as a third selection rule. 
Moreover, following arguments in~\cite{Cremades:2003qj} the couplings $\xi, \gamma, \mu$ involving the field $A_1$ only are constrained by
its origin from an ${\cal N}=2$ vector multiplet. For vanishing superpotential contributions to $A_1$, i.e. $\xi=\gamma=\mu=\alpha=0$, 
the first F-term condition~(\protect\ref{Eq:F-term_1}) is trivially fulfilled for arbitrary $\langle A_1 \rangle$, whereas the second F-term condition~(\protect\ref{Eq:F-term_2}) has the non-trivial solution
$\langle A_2 \rangle = -2\lambda/\beta$. Such a {\it vev} recombines D6-branes $a$ with their orbifold images $(\theta^k a)$ thereby generating 
field theoretical couplings, e.g. to $A_1$ or quarks, that were previously forbidden by stringy selection rules. 
Moreover, it was argued in~\cite{Cvetic:2008mh,Angelantonj:2009yj} that a Polonyi term ($\xi \neq 0$) might be generated by D2-brane instantons.
We therefore present the field theoretical discussion of the superpotential in this section in a general form including several types of $A_1$ couplings.
}

The $F$-term conditions for a supersymmetric minimum in Minkowski space are given by
\begin{eqnarray}\label{Eq:F-term_1}
0 \stackrel{!}{=} -F_{1}&=&\frac{\partial W}{\partial A_{1}}=\xi \unity_{3\times 3} +2\gamma A_{1}^{T}+\mu A_{1}^{T}A_{1}^{T}+\alpha A_{2}^{T}A_{2}^{T}
, \\
0 \stackrel{!}{=} -F_{2}&=&\frac{\partial W}{\partial A_{2}}=2\lambda A_{2}^{T}+\beta A_{2}^{T}A_{2}^{T}
+\alpha\left(A_{1}^{T}A_{2}^{T}+A_{2}^{T}A_{1}^{T}\right)
,
\label{Eq:F-term_2}
\end{eqnarray}
and with the ansatz  $\langle A_{1} \rangle=a \unity_{3\times 3}$ and $\langle A_{2} \rangle=ab \unity_{3 \times 3}$ with $a,b\in\mathbb{R}$ in order to preserve the $SU(3)_a$ gauge symmetry, 
or in other words only {\it vev}s for the singlet fields in equation~(\ref{Eq:Decompose_Adjoints_UN}), the $F$-term conditions reduce to 
\begin{equation}\label{Eq:F-terms-A}
0=\xi+2\gamma a+(\mu+\alpha b^{2})a^{2}
\qquad\text{and}\qquad
0=2\lambda ab+(\beta b+2\alpha)a^{2}b.
\end{equation}

For $\xi=0$, the $F$-term conditions~(\ref{Eq:F-terms-A}) have three types of solutions:
\begin{enumerate}
\item
Both fields $A_1,A_2$ have zero {\it vev}s, i.e. $a=0$. In this case, all right-handed vector-like quarks remain massless. 
\item
$A_1$ receives a non-trivial {\it vev}, $a=-\frac{2\gamma}{\mu}$, while $A_2$ remains trivial, i.e. $b=0$. 
In this case, one  vector-like combination of right-handed quarks of each kind $W_i$, $Y_i$, $Z_i$ acquires a mass, while all others remain massless.
\item
Both fields $A_1$ and $A_2$ receive supersymmetry preserving {\it vev}s ,
\begin{equation}
a=\frac{-2\gamma}{\mu+\alpha b^{2}}=\frac{-2\lambda}{\beta b+2\alpha}
\qquad\text{and}\qquad
b=\frac{1}{2\lambda\alpha}\left(3\gamma\beta\pm\sqrt{\gamma^{2}\beta^{2}-4\lambda\alpha(\mu\lambda-2\alpha\gamma)}\right)
.
\end{equation}
In this case, all vector-like left- and right-handed quarks acquire masses.
\end{enumerate}

For $\xi \neq 0$, two solutions can be distinguished:
\begin{enumerate}
\item
$A_2$ does not receive a {\it vev}, but $A_1$ does with $a=\frac{1}{\mu}\left(-\gamma\pm\sqrt{\gamma^{2}-\mu\xi}\right)$.
Again, only  one vector-like combination of each kind $W_i, Y_i, Z_i$ acquires  a mass, while all other right-handed vector-like quark pairs remain massless.
\item
Both $A_1$ and $A_2$ receive {\it vev}s,
\begin{equation}
a=\frac{-2\lambda}{\beta b+2\alpha}
\qquad\text{and}\qquad
b=\frac{2\beta(\xi\alpha-\lambda\gamma)\pm2\lambda\sqrt{\gamma^{2}\beta^{2}-4\lambda\alpha(\mu\lambda-2\alpha\gamma)-\xi(4\alpha^{3}+\beta^{2}\mu)}}{\xi\beta^{2}+4\alpha\lambda^{2}}
.
\end{equation}
In this case, all  right- and left-handed vector-like quark pairs receive masses.
\end{enumerate}

Since the superpotential~(\ref{SuperpotentialAdjoint}) contains cubic terms, a non-trivial {\it vev} of the singlet term inside  the adjoint $A_1$ of $U(3)_a$ renders 
the adjoint multiplet $\tilde{A}_{1}=({\bf 8}_{\Adj},\1)_0$ of $SU(3)_a \subset U(3)_a$ massive for $\mu \neq 0$. More explicitly, if we insert $A_{1}=a \unity_{3 \times 3} +\tilde{A}_{1}$ and 
$A_{2}=ab \unity_{3 \times 3}+\tilde{A}_{2}$ in the superpotential (\ref{SuperpotentialAdjoint}), we find 
\begin{equation}
\begin{aligned}
W=&(\gamma+\mu a)\text{Tr}(\tilde{A}_{1}^{2})+(\lambda+\alpha a+\beta ab)\text{Tr}(\tilde{A}_{2}^{2})+2\alpha ab\text{Tr}(\tilde{A}_{1}\tilde{A}_{2})
+ \ldots 
.
\end{aligned}
\end{equation}
The mass-terms for $\tilde{A}_{1}$ and $\tilde{A}_{2}$ are thus given by the mass matrix
\begin{equation}
M=\begin{pmatrix}\gamma+\mu \, a & \alpha \, ab \\ \alpha \, ab & \lambda+\alpha \, a+\beta \, ab\end{pmatrix}
 =\begin{pmatrix}\gamma+\mu \, a & \alpha \, ab \\ \alpha \, ab & \frac{1}{2}\beta \, ab\end{pmatrix}
,
\end{equation}
where in the last equality, the $F$-term condition (\ref{Eq:F-terms-A}) was used. Both adjoints $\tilde{A}_i$ of $SU(3)_a$ on
the first line of the `non-chiral' spectrum in equation~(\ref{FullSpectrum2}) are therefore expected to receive mass terms for $ab \neq 0$ 
unless the prefactors $\alpha, \beta \ldots$ in the superpotential take very special values, which could only be determined
in an elaborate string theoretic computation of $n$-point correlators on the $T^6/\Z_6'$ orbifold.

%%%%%%%%%%%%%%%%%%%%%%%%%%%%%%%%%%%%%%%%
\subsubsection{Comments on vacuum expectation values in the adjoints of $U(2)_b$}\label{Sss:Adjoints_U2b}

The `non-chiral' spectrum in equation~(\ref{FullSpectrum2}) contains ten multiplets in the adjoint representation $B_i$ of $U(2)_b$,
for which a complete list of localisations is given in table~\ref{tab:local_Adj_U2}. 
The computation of mass terms for the adjoints $\tilde{B}_i=(\1,\3_{\Adj})_0$ of $SU(2)_b$ under the decomposition
in equation~(\ref{Eq:Decompose_Adjoints_UN})
along $D$- and $F$-flat directions  is completely analogous to the one for the adjoints $\tilde{A}_i$ of $SU(3)_a$
in section~\ref{Sss:Adjoints_U3a} with {\it vev}s for the singlets inside $B_i$. 
The existence of nine adjoint representations $B_{i,i \in \{2 \ldots 10\}}$ at $b(\theta^k b)_{k=1,2}$ intersections leads to a non-trivial pattern of 
diagonal interaction terms at one point on the six-torus plus area suppressed mixings of three types (with $i \in \{1,2,3\}$), 
\begin{equation}
\begin{aligned}
W_{(B_{1+i})^m}, \quad 
W_{(B_{4+i})^n (B_{7+i})^p} & \sim {\cal O}(1)
\quad \text{ for } \quad
m,n+p = 2,3
,
\\ 
W_{B_2B_3B_4}, \quad 
W_{B_{5+3m_1}B_{6+3m_2}B_{7+3m_3}} & \sim {\cal O}(e^{-\mathcal{A}_{(2)}/6})
\quad \text{ for } \quad 
m_1,m_2,m_3 \in \{0,1\}
,
\\
W_{B_{1+i} (B_{4+i})^m (B_{7+i})^{2-m}}& \sim {\cal O}(e^{-\mathcal{A}_{(1)}/24})
\quad \text{ for } \quad
m \in \{0,1,2\}
,
 \\
W_{B_{1+i} B_{4+3m_1+j} B_{4+3m_2+k} }& \sim {\cal O}(e^{-\mathcal{A}_{(1)}/24-\mathcal{A}_{(2)}/6})
\quad \text{ for } \quad
m_1,m_2 \in \{0,1\}
,
\\
& \quad \text{ and } \quad 
(i,j,k) \text{ cyclic permutations of } (1,2,3)
.
\end{aligned}
\end{equation}
In addition, all kinds of couplings of the adjoint $B_1$ from the $bb$ sector, ${\rm Tr}(B_1^k)_{k\in \{1,2,3\}}$ and 
 ${\rm Tr}(B_1 B_{1+i}^2)$ and ${\rm Tr}(B_1(B_{4+i})^n (B_{7+i})^{2-n})_{n \in \{0,1,2\}}$,  contribute to the superpotential without area suppression
 in analogy to equation~(\ref{SuperpotentialAdjoints}).
This rich structure is expected to be able to provide supersymmetric mass terms for all ten
adjoints $\tilde{B}_i = (\1,\3_{\Adj})_0$ of $SU(2)_b$ on the first line 
of the `non-chiral' spectrum in equation~(\ref{FullSpectrum2}) simultaneously to those of the symmetrics $(b_i,\ov{b}_i)$ discussed in
section~\ref{Sss:VectorSymmetrics} with leading couplings to the $B_i$ listed in table~\ref{tab:Sym+Adj_U2b} of appendix~\ref{App:B}.
As for the adjoints of $SU(3)_a$, a more precise discussion of the mass terms requires the derivation of the correct prefactors in the
superpotential via sophisticated string theoretic methods such as $n$-point correlation functions on the $T^6/\Z_6'$ background.

%%%%%%%%%%%%%%%%%%%%%%%%%%%%%%%%%%%%%%%%%%%%%%%%%%%%%%%%%%%%%%%%%%%%%%%%%%%%%%
\subsection{Vacuum expectation values of symmetric representations of $U(2)_b$}\label{Vacuum expectation value of (anti)symmetric representations}

The symmetric representation $S$ of $U(N)$ transforms as $S'=USU^{T}$ under gauge transformations, and by considering infinitesimal
transformations $U=\unity_{N \times N} +i\alpha$ with $\alpha \equiv \sum \alpha_a t^a$, where $t^a$ are the generators of $U(N)$,  we can derive the generators 
$T^{a}_{[ij][kl]}=t^{a}_{ik}\delta_{jl}+t^{a}_{jl}\delta_{ik}$ of the symmetric representation. This leads to the $D$-term 
\begin{equation}\label{Eq:D-term-SUN-Sym}
D^{a}=-gS_{[ij]}T^{a}_{[ij][kl]}S_{[kl]}^{*} =-2g\text{Tr}(S^{\dagger}t^{a}S).
\end{equation}
Using 
\begin{equation}\label{Eq:tt_SUN}
\sum_{a}t^{a}_{ij}t^{a}_{kl}=\frac{1}{2}\left(\delta_{il}\delta_{jk}-\frac{1}{N}\delta_{ij}\delta_{kl}\right)
\quad \text{ for } \quad
SU(N),
\end{equation}
 the contribution of a symmetric representation to the $D$-terms  scalar potential is reduced to
\begin{eqnarray}
V_{D}
&=&\frac{1}{2}\sum_{a}D^{a}D^{a}
=g^{2}\left(\text{Tr}(S^{\dagger}SS^{\dagger}S)-\frac{1}{2}\text{Tr}(S^{\dagger}S)\text{Tr}(S^{\dagger}S)\right)
\geq 0
\quad 
\text{ for } SU(2).
\end{eqnarray}
%for $SU(2)$.
A flat direction, $V_D=0$ of this $D$-term contribution is given by the parameterisation
\begin{equation}
S=e^{i\alpha} \, \begin{pmatrix}
m & iq \\ iq & m
\end{pmatrix}
\qquad\Rightarrow\qquad
S^{\dagger}S=(m^{2}+q^{2}) \, \unity_{2 \times 2}
,
\label{FlatDirections}
\end{equation}
for which also $D^a=0$, as can be explicitly computed using the Pauli matrices for the generators $t^a$ of $SU(2)_b$.

At this point it is important to notice, that for the full group $U(2)_b$ equation~(\ref{Eq:tt_SUN}) is replaced by 
$\sum_{a}t^{a}_{ij}t^{a}_{kl}=\frac{1}{2}\delta_{il}\delta_{jk}$, and a massless $U(1)_b \subset U(2)_b$
does not allow for flat directions, $V_{D}=g^{2}\text{Tr}(S^{\dagger}SS^{\dagger}S) > 0$
for $S \neq 0$. However, the Abelian group $U(1)_b$ in the Standard Model with hidden $USp(6)_h$ on $T^6/\Z_6'$
is anomalous and acquires a string scale mass via the generalised Green-Schwarz mechanism justifying our ansatz for the $D$-terms of $SU(2)_b$ only.

In section~\ref{Sss:Leptons}, we argued that three symmetric representations $\ov{b}_i$ are needed to render all vector-like lepton pairs massive.
Assuming {\it vev}s of the form~(\ref{FlatDirections}) for each representation clearly corresponds to a flat direction, $V_D=0$, since each
contribution to the $D$-terms, $D^a=-2g \sum_{i} \text{Tr}(S_{i}^{\dagger}t^{a}S_{i}) =0$, vanishes.

Similar to the discussion of {\it vev}s for the adjoints $B_i$ of $U(2)_b$, the $F$-term contributions from the symmetric representations
contain all possible interaction terms. Since the prefactors of the couplings are not known, we assume that a non-trivial 
vacuum with {\it vev}s of some symmetric representations of  $U(2)_b$ exists.
In contrast to the discussion of {\it vev}s for adjoints of $U(2)_b$ in section~\ref{Ss:Adjoint-vevs}, the symmetric representation 
is, however, irreducible and any {\it vev} breaks the $SU(2)_b$ gauge symmetry, as can be seen for example from the covariant derivative,
\begin{align}
D_{\mu}S&=\partial_{\mu}S-ig(W_{\mu}S+SW_{\mu}^{T})
,
\end{align}
where $W_{\mu}$ is the gauge potential of $SU(2)_b$. For example,  $\langle S \rangle = \begin{pmatrix}0&\mu\\\mu&0\end{pmatrix}$
generates the mass term $8g^{2}|\mu|^{2}W_{\mu}^{+}W_{\mu}^{-}$ with $W_{\mu}^{\pm}=\frac{1}{2}(W_{\mu}^{1}\pm iW_{\mu}^{2})$ the standard definition of 
the charged vector-bosons of the weak gauge symmetry. In contrast to the Higgs mechanism $SU(2) \times U(1)_Y \to U(1)_{\text{el-mag}}$
of the Standard Model, only two out of three gauge bosons are rendered massive by the choice of some {\it vev}s inside symmetric representations. 
We therefore postulate  a step-wise breaking down to the electromagnetic gauge symmetry, 
\begin{equation*}
\begin{array}{lll}
SU(2)_b \times U(1)_Y \quad  \stackrel{\text{at } M_{\text{sym}} < M_{string}}{\longrightarrow} \quad 
& U(1)_{I_3} \times U(1)_Y  \quad \stackrel{\text{at } M_{\text{el-mag}} < M_{\text{sym}}}{\longrightarrow} \quad 
& U(1)_{\text{el-mag}}
,
\end{array}
\end{equation*}
with an intermediary scale $M_{\text{sym}}$ below which the third component of the weak isospin times the hyper charge, $U(1)_{I_3} \times U(1)_Y$,
remain massless. Such a mechanism is expected to affect the Weinberg angle $\theta_W$, which for our model is given
at $M_{string}$ by $\sin^2 \theta_W =0.65$~\cite{Gmeiner:2009fb}. Since the angle receives quantum corrections which depend on the energy scale,
matching data at the electro-weak scale requires an in-depth study of all participating mass scales. This goes clearly beyond the scope of this article.

The above discussion focussed on the symmetric representations of $U(2)_b$ which occur in the model. 
A different way of realising that the $D$-term conditions~(\ref{Eq:D-term-SUN-Sym}) are trivially fulfilled and the $SU(2)_b$ symmetry broken by 
an arbitrary {\it vev} relies on the fact that $SU(2)$ has exactly one three-dimensional representation, and therefore the symmetric representation
is equivalent to the adjoint representation.

In related models, it is possible that antisymmetric representations $(\1,\1_{\Anti})_0$ instead of symmetric matter of $U(2)_b$ couple to the vector-like 
lepton or Higgs pairs. In this case, it is again important for the $D$-term condition that $U(1)_b$ is anomalous and massive, but any {\it vev} will preserve 
the $SU(2)_b \subset U(2)_b$ gauge symmetry.

%%%%%%%%%%%%%%%%%%%%%%%%%%%%%%%%%%%%%%%%%%%%%%%%%%%%%%%%%%%%%%%%
\section{Couplings to the `hidden' $USp(6)_h$ sector}\label{S:Couplings_hidden}

Up to now, the discussion has focussed on supersymmetric mass terms for vector-like multiplets with only Standard Model charges.
The states with hidden sector charges in equation~(\ref{FullSpectrum3}) consist of antisymmetric matter $X_1,X_2$ of $USp(6)_h$ coupling
only gravitationally to the Standard Model plus vector-like pairs $\Pi^{\pm}$ and $\Omega^{\pm}_i$ with $SU(2)_b$ or $U(1)_Y$ charge, 
respectively, but no gauge interaction with $SU(3)_a \times U(1)_{B-L}$. 

$USp(2N)$ gauge groups are good candidates for supersymmetry breaking via gaugino condensation in a strongly coupled phase.  
To reach the strong coupling regime at energies below $M_{string}$, the charged matter states must acquire masses. 
Analogously to the pairs of microscopic ${\cal N}=1$ supersymmetric sectors in sections~\ref{Sss:Leptons} to~\ref{Sss:VectorSymmetrics}, 
the `hidden' sector fields with $SU(2)_b$ or $U(1)_Y$ charge have (at the level of triangular worldsheets) 
area suppressed three-point interactions with the antisymmetric representation $X_2$ of $USp(6)_h$ at $h(\theta^k h)_{k=1,2}$ intersections listed  
on the last two lines of table~\ref{Hidden-Sequences} in appendix~\ref{App:B}.
Assuming that the discussion for a {\it vev} of $X_2$ proceeds analogous to the discussion of the (anti)symmetrics of $U(2)_b$ in
section~\ref{Vacuum expectation value of (anti)symmetric representations}, all `hidden' sector fields are rendered massive.

The `hidden' sector matter fields are of interest also from a different point of view. Supposing that 
supersymmetry breaking is realised in the hidden sector, the fields with charges under $SU(2)_b \times U(1)_Y$ and $USp(6)_h$ 
might act as messenger fields, which transmit the supersymmetry breaking to the visible sector. In that sense, also the 
area-suppressed couplings $\Omega_2 \Pi^+ H_{6+i}$ with $i \in \{1,2,3\}$ on the first lines  of table~\ref{Hidden-Sequences} are of interest since 
in particular $i=1$  corresponds to the Higgs multiplet $H_7$ which participates in the non-suppressed quark Yukawa interactions of 
section~\ref{Sss:Quarks} and table~\ref{YukawaForBaryons} of appendix~\ref{App:B}.

%%%%%%%%%%%%%%%%%%%%%%%%%%%%%%%%%%%%%%%%%%%%%%%%%%%%%%%%%%%%%%%%
\section{Conclusions and Outlook}\label{S:Conclusions}

We exhaustively investigated the perturbative leading order three-point interactions of matter 
states in the Standard Model with hidden $USp(6)_h$ on intersecting D6-branes on $T^6/\Z_6'$ 
from~\cite{Gmeiner:2008xq,Gmeiner:2009fb}.
For the lepton Yukawa interactions, we found that the dominant terms are flavour diagonal and
involve one Higgs generation per lepton family, whereas the quark sector contains even at the leading order
mixings and two further Higgs generations. It will be interesting to perform a more detailed 
investigation of the rich flavour structure and Yukawa hierarchies
of both the lepton and quark sector for this model in the future.

Based on the same selection rules of charge cancellation and on the existence of closed triangular
worldsheets, we further found that (nearly) all vector-like charged states in the open 
string spectrum receive masses through Higgs-like couplings involving some fields with {\it vev}s: 
\begin{itemize}
\item
two displacements $\sigma^2_b, \sigma^2_c$ render all vector-like
exotic leptons on the last two lines of~(\ref{FullSpectrum2}) massive
while also creating Yukawa hierarchies within a given particle generation, 
\item
two {\it vev}s for the singlets inside the adjoints $A_1,A_2$ of $U(3)_a$
render all vector-like quarks on the second line of~(\ref{FullSpectrum2}) massive,
while also providing mass terms for the two adjoints of $SU(3)_a$ on the first line
of equation~(\ref{FullSpectrum2}),
\item
the vector-like symmetric representations of $SU(2)_b$ on the second line of~(\ref{FullSpectrum2})
and the adjoints of $SU(2)_b$ on the first line of~(\ref{FullSpectrum2}) receive masses
via couplings to adjoints of $U(2)_b$, where {\it vev}s of (at least) three singlets are needed,
\item
the `hidden' sector fields with $SU(2)_b \times U(1)_Y$ charge acquire masses if one antisymmetric
representation $X_2$ of $USp(6)_h$ acquires a {\it vev},
\item
the vector-like lepton pairs and six of the nine Higgs generations in~(\ref{FullSpectrum1})
receive masses via perturbative three-point couplings to six symmetric representations of $SU(2)_b$, 
for which a {\it vev} breaks the gauge group to the third component of the weak isopsin, $U(1)_{I_3}$.
The remaining three Higgs generations may acquire masses when also couplings to massive matter states
are taken into account.
\end{itemize}
In summary, (at least) 2+2+3+1=8 {\it vev}s are needed to provide masses for all non-chiral states 
with Standard Model charges in equations~(\ref{FullSpectrum2}) and~(\ref{FullSpectrum3}). 
The vector-like lepton pairs and (most) Higgs generations in equation~(\ref{FullSpectrum1}) acquire 
at the level of perturbative three-point interactions masses if six further {\it vev}s in the symmetric 
and its conjugate representation of $U(2)_b$ are switched on. The latter will, however break 
$SU(2)_b \to U(1)_{I_3}$, and a thorough discussion of step-wise breaking  the gauge symmetry
down to the electro-magnetic group at the electro-weak scale needs to be performed in the future.
Taking into account non-renormalisable higher order couplings in the superpotential, which include neutral 
closed string fields as well, might reduce the number of required {\it vev}s while simultaneously
addressing the question of moduli stabilisation.

The Standard Model sector on D6-branes $a,b,c,d$ for the model with hidden $\widehat{USp(2)}_{\hat{h}}$ in~\cite{Gmeiner:2008xq,Gmeiner:2009fb}
coincides with the one with hidden $USp(6)_h$ discussed here, therefore all Yukawa couplings and mass terms - except for the ones involving hidden sector fields
in section~\ref{S:Couplings_hidden} - presented in this article are identical for both known D6-brane examples of the Standard Model
on $T^6/\Z_6'$ with some hidden sector.

The present result of 8+6 {\it vev}s when considering only perturbative three-point interactions can be compared
to known results for the Standard Model on heterotic orbifolds such as on $T^6/\Z_6'$ 
in~\cite{Lebedev:2008un}, where non-renormalisable couplings up to the sixth order in the Standard Model 
singlet fields with {\it vev}s were taken into account. Also in the  heterotic $T^6/\Z_2 \times \Z_2$ case 
of~\cite{Blaszczyk:2009in}, a larger number, 44, of {\it vev}s  was required to render the vector-like 
matter states massive. This may be related to the fact that we did not discuss masses for singlet fields 
such as the $h_{21}$ complex structure and $h_{11}^-$ K\"ahler moduli which arise in the closed string sector.
In heterotic orbifolds, the geometric moduli at fixed points receive charges under the gauge groups and
lead to enhancements of the representations. They are therefore necessarily included in any discussion 
of charged vector-like states within orbifold compactifications of the heterotic string.

The investigation in this article relies on two basic selection rules of the open string 
perturbation theory, namely charge cancellation including the anomalous Abelian factors and the existence of closed
triangles. In a fully string theoretic computation, these selection rules might have to be supplemented
by additional `intrinsic' symmetry properties such as the underlying $\Z_2$ orbifold symmetry.
To finally decide on the existence of $n$-point interactions and determine their strength, it will be 
necessary to generalise the derivation of Yukawa couplings on the six-torus to $T^6/\Z_{2N}$
and $T^6/\Z_2 \times \Z_{2M}$ orbifolds e.g. using correlators along the 
lines of~\cite{Abel:2005qn,Berg:2005ja,Anastasopoulos:2011kr,Anastasopoulos:2011gn} and the 
recent computation of two-point functions on $T^6/\Z_6'$ in~\cite{Berg:2011ij}. 
Such a computation will involve the complete worldsheet instanton sum in contrast to the leading
terms with the smallest possible triangular worldsheets considered here, and it will settle the open question if the 
suppression factor by $1/v_i^{1/4}$  for D6-branes parallel along $T^2_{(i)}$ - for which to our knowledge there exist no results in 
the literature - is of physical relevance to the Yukawa couplings, or if it cancels among the non-holomorphic prefactor and the 
holomorphic worldsheet instanton sum. 

Last but not least, the non-holomorphic K\"ahler metrics which contribute to the prefactor of the physical Yukawa interactions
are only known classically and can in principle receive corrections at any order in perturbation theory and by all kinds
of non-perturbative effects. It will be interesting to see if they are indeed protected by `intrinsic' string theoretic symmetries
as suggested recently in~\cite{Berg:2011ij}. 

\vspace{-2mm}

%%%%%%%%%%%%%%%%%%%%%%%%%%%%%%%%%%%%%%%%%%%%%%%%%%%%%%%%%%%%%%%%
\subsection*{Acknowledgements}
G.H. thanks F.~Saueressig for discussions and F.~Marchesano for useful email correspondence. J.V. thanks W. Troost for discussions and advice.
\\
The work of G.\ H. \ is partially supported by the  ``Research Center Elementary Forces and Mathematical Foundations'' (EMG)
at the Johannes Gutenberg-Universit\"at Mainz.
The research of J.V. has been supported in part by the Belgian Federal Science Policy Office through the Interuniversity Attraction Pole 
IAP VI/11 and by FWO-Vlaanderen through project G011410N. 
\\
J.V. acknowledges the support of the department of physics of the KULeuven and the kind hospitality at JGU Mainz 
at various stages of this project, which is partly based on the master thesis~\cite{Vanhoof:2011}.

%%%%%%%%%%%%%%%%%%%%%%%%%%%%%%%%%%%%%%%%%%%%%%%%%%%%%%%%%%%%%%%%

\clearpage
\begin{appendix}

\section{Tables with localisations of massless matter states}\label{App:A}

This appendix displays for the first time all sector-wise localisations of matter states of the Standard Model with hidden 
$USp(6)_h$ on $T^6/\Z_6'$, for which the matter spectrum had been computed in~\cite{Gmeiner:2008xq,Gmeiner:2009fb}.
The corresponding complete K\"ahler metrics are also given in tables~\ref{IntersectionNumbers-1} and~\ref{IntersectionNumbers-2},
where those for strings with an endpoint on the stacks $a$ or $d$ have been already been computed in~\cite{Honecker:2011sm}.
The positions of the intersection points in each sector are given by the apexes in the tables of three-point functions in 
appendix~\ref{App:B}.
{\small 
%%%%%%%%%%%%%%%%%%%%%%%%%%%%%%%%%%%%%%%%%
\begin{sidewaystable}[ht!]
\begin{center}
\setlength{\tabcolsep}{13pt}
{\footnotesize
\begin{tabular}{|c||c||c|c|c||c|c|c|}\hline
\multicolumn{8}{|c|}{\text{\bf Multiplicities and K\"ahler metrics of bifundamental and adjoint matter}}
\\ \hline\hline
\begin{sideways} \!\!\!\!\! Particle \end{sideways}&
$xy$  &\!\!\!$\chi_{xy}$ or $\varphi_{xy}$\!\!\!&\!\!\!$\chi_{x(\theta y)}$ or $\varphi_{x(\theta y)}$\!\!\!&
\!\!\!$\chi_{x(\theta^2 y)}$ or $\varphi_{x(\theta^{2}y)}$ & $K_{xy}$  & $K_{x(\theta y)}$   & $K_{x(\theta^2 y)}$  
\\\hline
$A_i$ & $aa$ & 1 & \muc{2}{|c||}{1} & $f(S,U)\frac{1}{v_2}\sqrt{\frac{\pi \, r}{2}}$ & \muc{2}{|c|}{$f(S,U)\sqrt{\frac{2\pi \, r}{v_1v_2}}$} \\
$B_i$ & $bb$ & 1 & \muc{2}{|c||}{9} & $f(S,U)\sqrt{\frac{4\pi\sqrt{3}}{v_{1}v_{3}}}$ & \muc{2}{|c|}{$f(S,U)\sqrt{\frac{2\pi}{v_{1}v_{2}}\left(\frac{1}{r}+r\right)}$} \\
& $cc$ & 1 & \muc{2}{|c||}{3} & $f(S,U)\sqrt{\frac{4\pi\sqrt{3}}{v_{1}v_{3}}}$ & \muc{2}{|c|}{$f(S,U)\sqrt{\frac{2\pi}{rv_{1}v_{2}}}$} \\
& $dd$ & 1 & \muc{2}{|c||}{9} & $f(S,U)\sqrt{\frac{4\pi\sqrt{3}}{v_{1}v_{3}}}$ & \muc{2}{|c|}{$f(S,U)\sqrt{\frac{2\pi \, r}{v_{1}v_{2}}}$}
\\\hline
& $ab$  & $-$      & $-$             & $-$ & $-$ & $-$ & $-$  \\
$U_i$ & $ac$  & $2$      & $-$             & $1$ & $f(S,U)\sqrt{\frac{4\pi}{\sqrt{3}v_{2}v_{3}}}$ & $-$ & $f(S,U)\sqrt{\frac{10}{v_{1}v_{2}v_{3}}}$ \\
$Y_i$ & $ad$  & $|4|$    & $-1$            & $1$ & $f(S,U)\sqrt{\frac{2\pi \, r}{v_1v_2}}$ & $f(S,U)\sqrt{\frac{2\pi \, r}{v_1v_2}}$ & $f(S,U)\sqrt{\frac{2\pi \, r}{v_1v_2}}$ \\
$H_i$ & $bc$  & $6$      & $3$             & $|2|_m$ & $f(S,U)\sqrt{\frac{10}{v_{1}v_{2}v_{3}}}$ & $f(S,U)\sqrt{\frac{25}{2v_{1}v_{2}v_{3}}}$ & $f(S,U)\sqrt{\frac{4\pi\sqrt{3}}{v_{1}v_{3}}}$ \\
$L_i$ & $bd$  & $-$      & $-6$            & $|4|_m$ & $-$ & $f(S,U)\sqrt{\frac{8}{v_{1}v_{2}v_{3}}}$ & $f(S,U)\sqrt{\frac{4\pi\sqrt{3}}{v_{1}v_{3}}}$ \\
$E_i$ & $cd$  & $|2|_m$    & $3$             & $-$ & $f(S,U)\sqrt{\frac{4\pi\sqrt{3}}{v_{1}v_{3}}}$ & $f(S,U)\sqrt{\frac{10}{v_{1}v_{2}v_{3}}}$ & $-$  \\
\hline
\hspace{-3mm}$Q_i,\ov{Q}$\hspace{-3mm}   & $ab'$& $-2$     & $-2$            & $1$ & $f(S,U)\sqrt{\frac{10}{v_{1}v_{2}v_{3}}}$ & $f(S,U)\sqrt{\frac{10}{v_{1}v_{2}v_{3}}}$ & $f(S,U)\sqrt{\frac{25}{2v_{1}v_{2}v_{3}}}$ \\
$Z_i$ & $ad'$ & $-1$     & $|4|$           & $1$ & $f(S,U)\sqrt{\frac{2\pi \, r}{v_1v_2}}$ & $f(S,U)\sqrt{\frac{2\pi \, r}{v_1v_2}}$ & $f(S,U)\sqrt{\frac{2\pi \, r}{v_1v_2}}$ \\
$\ov{L}_i$ & $bd'$ & $|2|_m$    & $-$             & $-3$ & $f(S,U)\sqrt{\frac{4\pi\sqrt{3}}{v_{1}v_{3}}}$ & $-$ & $f(S,U)\sqrt{\frac{8}{v_{1}v_{2}v_{3}}}$  \\
\hline
& $ah$  & $-$  & $-$         & $-$ & $-$ & $-$ & $-$  \\
$\Pi^{\pm}$ & $bh$  & $-1$ & $1$         & $-$ & $f(S,U)\sqrt{\frac{10}{v_{1}v_{2}v_{3}}}$ & $f(S,U)\sqrt{\frac{25}{2v_{1}v_{2}v_{3}}}$ & $-$  \\
$\Omega_i^{\pm}$& $ch$  & $-$  & $-1$        & $1$ & $-$ & $f(S,U)\sqrt{\frac{10}{v_{1}v_{2}v_{3}}}$ & $f(S,U)\sqrt{\frac{10}{v_{1}v_{2}v_{3}}}$  \\
& $dh$  & $-$  & $-$         & $-$ & $-$ & $-$ & $-$  \\
\hline
\end{tabular}
}
\end{center}
\caption{Multiplicities and K\"ahler metrics for matter in the adjoint and bifundamental matter for the Standard Model on D6-branes.
Whenever the matter in a given bifundamental sector is chiral, $\chi$ counts the number of states with the sign representing their chirality.
If on the other hand, matter in a given bifundamental sector is non-chiral the total amount $\varphi$ in terms of an absolute value $|\cdot|$.
The special case of a microscopic ${\cal N}=2$ origin with D6-branes parallel along $T^2_{(2)}$ is labelled by $|\cdot|_m$ where the index $m$
refers to the possibility of generating mass terms by relative parallel D6-brane displacements. \newline
Supersymmetry requires $r=\frac{1}{\sqrt{3}}$ for the complex structure of $T^2_{(3)}$. $v_i$ denote the two-torus $T^2_{(i)}$ volumes.
}
\label{IntersectionNumbers-1}
\end{sidewaystable}
%%%%%%%%%%%%%%%%%%%
}
%
%%%%%%%%%%%%%%%%%%%%%%%%%%%%%%%%%%%%%%%%%
\begin{sidewaystable}[ht!]
\begin{center}
\setlength{\tabcolsep}{10pt}
{\footnotesize
\begin{tabular}{|c||c||c|c||c|c||c|c||c|c|c|}\hline
\multicolumn{11}{|c|}{\text{\bf Multiplicities and K\"ahler metrics of (anti)symmetric matter}}
\\ \hline\hline
\begin{sideways} \!\!\!\!\! Particle \end{sideways}&$x$  &\begin{sideways}$\left(\chi/\varphi\right)^{\Anti}_{xx'}$\end{sideways}
&\begin{sideways} $\left(\chi/\varphi\right)^{\Sym}_{xx'}$ \end{sideways}
 &\begin{sideways} $\left(\chi/\varphi\right)^{\Anti}_{x(\theta x')}$\end{sideways}
&\begin{sideways} $\left(\chi/\varphi\right)^{\Sym}_{x(\theta x')}$\end{sideways}
 &\begin{sideways} $\left(\chi/\varphi\right)^{\Anti}_{x(\theta^2 x')}$\end{sideways}
&\begin{sideways} $\left(\chi/\varphi\right)^{\Sym}_{x(\theta^2 x')}$\end{sideways}
& \!\!$K_{xx'}$\!\!  & \!\!$K_{x(\theta x')}$\!\!   & \!\!$K_{x(\theta^2 x')}$\!\!  
\\\hline
$W_i$ & $a$ & -1 & - & $|4|$ & - & 1 & - &$f(S,U) \sqrt{\frac{2\pi \, r}{v_1v_2}}$ &$f(S,U) \sqrt{\frac{4\pi}{\sqrt{3}v_1v_3}}$ 
& $f(S,U) \sqrt{\frac{2\pi \, r}{v_1v_2}}$
\\\hline
$b_i, \ov{b}_i$ & $b$ & - & 6 & - & -6 & $|8|_m$ & - & $f(S,U)\sqrt{\frac{8}{v_{1}v_{2}v_{3}}}$ & $f(S,U)\sqrt{\frac{4\pi\sqrt{3}}{v_{2}v_{3}}}$ & $f(S,U)\sqrt{\frac{4\pi\sqrt{3}}{v_{1}v_{3}}}$
\\\hline
& $c$ & $(\frac{3}{2} \to \emptyset)$ & - & - & $1 \to |2|_m$ & $(\frac{3}{2} \to \emptyset)$ & - &  $f(S,U)\sqrt{\frac{2\pi}{rv_{1}v_{2}}}$ & $f(S,U)\sqrt{\frac{4\pi\sqrt{3}}{v_{1}v_{3}}}$ & $f(S,U)\sqrt{\frac{2\pi}{rv_{1}v_{2}}}$
\\\hline
& $d$ & (9) & - & ($|4_m|$) & - & (-9) & - & $-$ & $-$ & $-$
\\\hline
$X_i$ & $h$ & $\frac{1}{2}$ & - & 1 & - & $\frac{1}{2}$ & - & $f(S,U)\sqrt{\frac{2\pi \, r}{v_{1}v_{2}}}$ & $f(S,U)\sqrt{\frac{4\pi}{\sqrt{3}v_{1}v_{3}}}$ & $f(S,U)\sqrt{\frac{2\pi \, r}{v_{1}v_{2}}}$
\\\hline
\end{tabular}
}
\end{center}
\caption{Multiplicities and K\"ahler metrics for symmetric and antisymmetric matter of the Standard Model on D6-branes.
The stacks $c$ and $h$ carry the gauge groups $USp(2)_c$ and $USp(6)_h$. For the former the decomposition under $USp(2)_c \to U(1)_c$
is given. The formal multiplicities of non-existent antisymmetric matter under $U(1)_d$ are for the sake of completeness given in parenthesis. 
For more details on the notation see the caption of table~\protect\ref{IntersectionNumbers-1}.}
\label{IntersectionNumbers-2}
\end{sidewaystable}
%%%%%%%%%%%%%%%%%%%

\clearpage

\section{Tables with suppression factors of three-point interactions}\label{App:B}

All relevant allowed three-point interactions are listed in this appendix. An allowed three-point coupling corresponds to a cyclically ordered sequence $[xy]+[yz]+[zx]\leadsto[x,y,z]$. The oriented strings then form a closed triangle, and the corresponding interaction term is gauge invariant
and produced by worldsheet instantons sweeping the area.
%%%%%%%%%%%%%%%%%%

{\footnotesize
%%%%%%%%%%%%%%%%%%
\begin{table}[ht!]
{\footnotesize
\begin{center}
\setlength{\tabcolsep}{4pt}
\begin{tabular}{|c||c|c|c|c|}
\hline
\muc{5}{|c|}{\text{\bf Lepton Yukawa couplings before breaking $USp(2)_c \to U(1)_c$}}
\\\hline\hline
\begin{tabular}{c}sequence\\$[x,y,z]$\\\end{tabular} & \begin{tabular}{c}enclosed\\area\\\end{tabular} & \begin{tabular}{c}triangle\\(or point)\\\end{tabular} & coupling & \begin{tabular}{c}K\"ahler factor\\$(K_{xy}K_{yz}K_{zx})^{-1/2}$\\\end{tabular} \\ \hline\hline
$[c,(\theta d),b]$ & $0$          & $\{4,i,3\}$                  & $E_{i}L_{i}H_{i}$, $i \in \{1,2,3\}$ & $\left(\frac{v_1v_2v_3}{f(S,U)^2}\right)^{3/4}\frac{1}{2(50)^{\frac{1}{4}}}$ \\                               \cline{2-4}
                               & $\frac{1}{12}\mathcal{A}_{(1)}$ & $\{[(R,R'),4,5],i,3\}$ & $E_{i}L_{3+i}H_{3+i}$, $i \in \{1,2,3\}$  &\\
                               \cline{2-4}
                               & $\frac{1}{6}\mathcal{A}_{(2)}$ & $\{4,[1,2,3],3\}$       & $\begin{array}{c} E_{i}L_{j}H_{k}
\text{ with } (i,j,k) \\ \text{permutations of } (1,2,3)
\end{array} $ &\\\cline{2-4}
                               & $\frac{1}{12}\mathcal{A}_{(1)}+\frac{1}{6}\mathcal{A}_{(2)}$ & $\{[(R,R'),4,5],[1,2,3],3\}$ & 
$\begin{array}{c} E_{i}L_{j}H_{k} \text{ with }\\ (i,j,k)=(1,\underline{5,6}),(2,\underline{4,6}),(3,\underline{4,5}) \end{array}$ &\\ \cline{2-4}
                              & $\frac{1}{3}\mathcal{A}_{(1)}$ & $\{[4,(R,R'),4],i,3\}$ & $E_{i}L_{3+i}H_{i}$, $i \in \{1,2,3\}$  &\\ \cline{2-4}
                              & $\frac{1}{3}\mathcal{A}_{(1)}+\frac{1}{6}\mathcal{A}_{(2)}$ & $\{[(R,R'),4,5],[1,2,3],3\}$ & 
$\begin{array}{c} E_{i}L_{j}H_{k-3} \text{ with }\\ (i,j,k)=(1,\underline{5,6}),(2,\underline{4,6}),(3,\underline{4,5}) \end{array}$ &\\ 
\hline
\end{tabular}
\end{center}
}
\caption{Three-point couplings between leptons and Higgses which are allowed by charge selection rules and the existence of triangular worldsheets.
The sizes of triangular worldsheets are given before breaking the right-symmetric group $USp(2)_c \to U(1)_c$ by a continuous displacement $\sigma^2_c$
as discussed in section~\protect\ref{Sss:Breaking}.}
\label{YukawaForLeptons}
\end{table}
%%%%%%%%%%%%%%%%%%
}
%%%%%%%%%%%%%%%%%
\begin{table}[ht!]
{\footnotesize
\begin{center}
\setlength{\tabcolsep}{4pt}
\begin{tabular}{|c||c|c|c|c|}
\hline
\muc{5}{|c|}{\text{\bf Quark Yukawa couplings before breaking $USp(2)_c \to U(1)_c$}}
\\\hline\hline
\begin{tabular}{c}sequence\\$[x,y,z]$\\\end{tabular} & \begin{tabular}{c}enclosed\\area\\\end{tabular} & \begin{tabular}{c}triangle\\(or point)\\\end{tabular} & coupling & \begin{tabular}{c}K\"ahler f
actor\\$(K_{xy}K_{yz}K_{zx})^{-1/2}$\\\end{tabular} \\ \hline\hline
$[b',a,(\theta^{2}c)]$           & $0$    & $\{4,1,3\}$            & $Q_{1}U_{3}H_{1}$ & $\left(\frac{v_1v_2v_3}{f(S,U)^2}\right)^{3/4}\frac{1}{(10)^{\frac{3}{4}}}$  \\ \cline{2-4}
                                                     & $\frac{1}{12}\mathcal{A}_{(2)}$                                                   & $\{4,[(P,P'),1,i],3\}$ & $Q_{2}U_{3}H_{i}$, $i\in \{2,3\}$   &\\ \cline{2-4}
                                                     & $\frac{1}{4}\mathcal{A}_{(1)}$    & $\{[4,6,4],1,3\}$            & $Q_{1}U_{3}H_{4}$ & \\ \cline{2-4}
                                                     & $\frac{1}{4}\mathcal{A}_{(1)}+\frac{1}{12}\mathcal{A}_{(2)}$                                                       & $\{[4,6,4],[(P,P'),1,i],3\}$ & $Q_{2}U_{3}H_{3+i}$ , $i\in \{2,3\}$  &\\ \hline
$[(\theta b'),a,(\theta^{2}c)]$ & $0$    & $\{4,1,3\}$            & $Q_{4}U_{3}H_{7}$ & $\left(\frac{v_1v_2v_3}{f(S,U)^2}\right)^{3/4}\frac{1}{5(2)^{\frac{1}{4}}}$  \\ \cline{2-4}
                                              & $\frac{1}{6}\mathcal{A}_{(2)}$    & $\{4,[1,i,1],3\}$            & $Q_{4}U_{3}H_{6+i}$, $i\in \{2,3\}$  & \\
                                              \cline{2-4}
                                              & $\frac{1}{4}\mathcal{A}_{(1)}$    & $\{[4,5,4],1,3\}$            & $Q_{3}U_{3}H_{7}$ &  \\ \cline{2-4}
                                              & $\frac{1}{4}\mathcal{A}_{(1)}+\frac{1}{6}\mathcal{A}_{(2)}$    & $\{[4,5,4],[1,i,1],3\}$            & $Q_{3}U_{3}H_{6+i}$, $i\in \{2,3\}$   & \\
                                              \hline
$[(\theta b'),a,c]$                     & $0$    & $\{5,1,3\}$            & $Q_{3}U_{1}H_{1}$ & $\left(\frac{v_1v_2v_3}{f(S,U)^2}\right)^{3/4}\frac{(3)^{\frac{1}{8}}}{2(25\pi v_{1})^{\frac{1}{4}}}$   
\\
                                                     &        & $\{4,1,3\}$            & $Q_{4}U_{1}H_{4}$   &\\ \cline{2-4}
                                                     & $\frac{1}{12}\mathcal{A}_{(2)}$ & $\{5,[(P,P'),1,i],3\}$ & $Q_{3}U_{2}H_{5-i}$, $i\in  \{2,3\}$  &\\
                                                     &        & $\{4,[(P,P'),1,i],3\}$ & $Q_{4}U_{2}H_{8-i}$, $i\in  \{2,3\}$   &\\ \hline
\end{tabular}
\end{center}
}
\caption{Three-point couplings between quarks and Higgses before breaking the right-symmetric group.}
\label{YukawaForBaryons}
\end{table}
%%%%%%%%%%%%%%%%%%

\vspace{-5mm}

%%%%%%%%%%%%%%%%%%
\begin{table}[ht!]
{\footnotesize
\begin{center}\hspace{-8mm}
\setlength{\tabcolsep}{2pt}
\begin{tabular}{|c||c|c|c|c|}
\hline
\muc{5}{|c|}{\text{\bf Couplings between leptons or Higgses and symmetric matter}}
\\\hline\hline
\begin{tabular}{c}sequence\\$[x,y,z]$\\\end{tabular} & \begin{tabular}{c}enclosed\\area\\\end{tabular} 
& \begin{tabular}{c}triangle\\(or point)\\\end{tabular} & coupling & \begin{tabular}{c}K\"ahler factor\\$(K_{xy}K_{yz}K_{zx})^{-1/2}$\\\end{tabular} 
\\ \hline\hline
$[b',(\theta d),b]$ & $0$   & $\{(R,R'),i,3\}$       & $\ov{L}_{i}b_{i}L_{3+i}$, $i\in  \{1,2,3\}$ 
& $\left(\frac{v_1v_2v_3}{f(S,U)^2}\right)^{3/4}\!\!\!\frac{1}{4(2)^{\frac{1}{4}}}$ \\ \cline{2-4}
                    & $\frac{1}{6}\mathcal{A}_{(2)}$ & $\{(R,R'),[1,2,3],3\}$ & $\begin{array}{c} 
\ov{L}_{i}b_{j}L_{k} \text{ with } (i,j,k)=\\(\underline{2,3},4),(\underline{1,3},5),(\underline{1,2},6) \end{array}$&  \\ \cline{2-4}
                               & $\frac{1}{4}\mathcal{A}_{(3)}$ & $\{(R,R'),i,[3,1,3]\}$ & $\ov{L}_{i}b_{3+i}L_{3+i}$, $i\in  \{1,2,3\}$ & \\ \cline{2-4}
                               & $\frac{1}{6}\mathcal{A}_{(2)}+\frac{1}{4}\mathcal{A}_{(3)}$ & $\{(R,R'),[1,2,3],[3,1,3]\}$       & $\begin{array}{c} 
\ov{L}_{i}b_{3+j}L_{k} \text{ with } (i,j,k)=\\(\underline{2,3},4),(\underline{1,3},5),(\underline{1,2},6) \end{array}$ &
\\ \cline{2-4}
& $\frac{1}{6}\mathcal{A}_{(1)}$   & $\{[R,R',4],i,3\}$       & $\ov{L}_{i}b_{i}L_{i}$, $i\in  \{1,2,3\}$ &  \\ \cline{2-4}
                    & $\frac{1}{6}\mathcal{A}_{(1)}+\frac{1}{6}\mathcal{A}_{(2)}$ & $\{[R,R',4],[1,2,3],3\}$ & $\begin{array}{c} 
\ov{L}_{i}b_{j}L_{k} \text{ with } (i,j,k)=\\(\underline{2,3},1),(\underline{1,3},2),(\underline{1,2},3) \end{array}$&  \\ \cline{2-4}
                               & $\frac{1}{6}\mathcal{A}_{(1)}+\frac{1}{4}\mathcal{A}_{(3)}$ & $\{[R,R',4],i,[3,1,3]\}$ & $\ov{L}_{i}b_{3+i}L_{i}$, $i\in  \{1,2,3\}$ & \\ \cline{2-4}
                               & $\frac{1}{6} \mathcal{A}_{(1)}+\frac{1}{6}\mathcal{A}_{(2)} +\frac{1}{4}\mathcal{A}_{(3)}$ & $\{[R,R',4],[1,2,3],[3,1,3]\}$       & $\begin{array}{c} 
\ov{L}_{i}b_{3+j}L_{k} \text{ with } (i,j,k)=\\(\underline{2,3},1),(\underline{1,3},2),(\underline{1,2},3) \end{array}$ &
\\ \hline\hline
%%%%%%%%%%%%%%%%%%%%%%%%
$[c,(\theta b'),b]$ & $0$          & $\{4,i,3\}$                  & $\begin{array}{c} H_{j}\ov{b}_{i}H_{i} \text{ with } (i,j)=\\(1,4),(2,6),(3,5)\end{array}$ & $\left(\frac{v_1v_2v_3}{f(S,U)^2}\right)^{3/4}\frac{1}{2(25\pi\sqrt{3}v_{1})^{\frac{1}{4}}}$ \\ \cline{3-4}
                               &                  & $\{5,i,3\}$                  & $\begin{array}{c} H_{j}\ov{b}_{i}H_{3+i} \text{ with } (i,j)=\\(1,1),(2,3),(3,2)\end{array}$&\\ \cline{2-4}
                               & $\frac{1}{6}\mathcal{A}_{(2)}$ & $\{4,[1,2,3],3\}$       & $\begin{array}{c} H_{i}\bar{b}_{j}H_{k} \text{ with } (i,j,k)=\\(4,\underline{2,3}),(5,\underline{1,2}),(6,\underline{1,3}) 
                               \end{array}$ &\\ \cline{3-4}
                               &                 & $\{5,[1,2,3],3\}$ &
$\begin{array}{c} H_{i}\bar{b}_{j}H_{k} \text{ with } (i,j,k)=\\(\underline{2,3},4),(\underline{1,2},5),(\underline{1,3},6) \end{array}$ &
\\ \cline{2-4}
                               & $\frac{1}{4}\mathcal{A}_{(3)}$ & $\{4,i,[3,1,3]\}$ & $\begin{array}{c} H_{j}\ov{b}_{3+i}H_{i} \text{ with } (i,j)=\\(1,4),(2,6),(3,5) \end{array}$ & \\ \cline{3-4}
                               &   & $\{5,i,[3,1,3]\}$ & $\begin{array}{c} H_{j}\bar{b}_{i}H_{i} \text{ with } (i,j)=\\(4,1),(5,3),(6,2) \end{array}$ & \\ \cline{2-4}
                              & $\frac{1}{6}\mathcal{A}_{(2)}+\frac{1}{4}\mathcal{A}_{(3)}$ & $\{4,[1,2,3],[3,1,3]\}$       & $\begin{array}{c} H_{i}\ov{b}_{3+j}H_{k} \text{ with } (i,j,k)=\\(4,\underline{2,3}),
                              (5,\underline{1,2}),(6,\underline{1,3}) \end{array}$ &\\ \cline{3-4}
                               &                 & $\{5,[1,2,3],[3,1,3]\}$ &
$\begin{array}{c} H_{i}\ov{b}_{3+j}H_{k} \text{ with } (i,j,k)=\\(\underline{2,3},4),(\underline{1,2},5),(\underline{1,3},6) \end{array}$ &
\\\hline
\end{tabular}
\end{center}
}
\caption{Three-point couplings between the symmetrics $\ov{b}_i$ of $U(2)_b$ and the leptonic $L_j$ and $\ov{L}_j$ fields
as well as the symmetrics $b_i$ and the Higgs fields $H_j$.}
\label{LLbarSym}
\end{table}
%%%%%%%%%%%%%%%%%%

%
%%%%%%%%%%%%%%%%%%
\begin{table}[ht!]
{\footnotesize
\begin{center}
\setlength{\tabcolsep}{4pt}
\begin{tabular}{|c||c|c|c|c|}
\hline
\muc{5}{|c|}{\text{\bf Couplings between quarks and adjoint matter}}
\\\hline\hline
\begin{tabular}{c}sequence\\$[x,y,z]$\\\end{tabular} & \begin{tabular}{c}enclosed\\area\\\end{tabular} & \begin{tabular}{c}triangle\\(or point)\\\end{tabular} 
& coupling & \begin{tabular}{c}K\"ahler factor\\$(K_{xy}K_{yz}K_{zx})^{-1/2}$\\\end{tabular} \\ \hline\hline
$[(\theta a),b',a]$                     & $\frac{1}{8}\mathcal{A}_{(1)}$ & $\{[4,5,6],1,3\}$ & $\ov{Q}Q_{1}A_{2}$ 
& $\left(\frac{v_1v_2v_3}{f(S,U)^2}\right)^{3/4} \frac{1}{(250\pi\,rv_{3})^{\frac{1}{4}}}$ 
\\ \cline{2-4}
                            & $\frac{1}{8}\mathcal{A}_{(1)}+\frac{1}{4}\mathcal{A}_{(2)}$ & $\{[4,5,6],[1,P,1],3\}$                  & $\ov{Q}Q_{2}A_{2}$ & \\ \hline
$[(\theta^{2}a),(\theta b'),a]$ & $0$ & $\{4,1,3\}$ & $\ov{Q}Q_{4}A_{2}$ &
 $\left(\frac{v_1v_2v_3}{f(S,U)^2}\right)^{3/4}\frac{1}{(250\pi\,rv_{3})^{\frac{1}{4}}}$ \\ \cline{2-4}
                            & $\frac{1}{4}\mathcal{A}_{(1)}$ & $\{[4,5,4],1,3\}$                  & $\ov{Q}Q_{3}A_{2}$ & \
\\ \hline\hline
$[(\theta a),b',(\theta b')]$ & $\frac{1}{24}\mathcal{A}_{(1)}$ & $\{[6,(R,R'),5],1,3\}$ & $\begin{array}{c} \ov{Q}B_{k}Q_{1} \\  \text{with } k=5,8
\end{array}$ & $\left(\frac{v_1v_2v_3}{f(S,U)^2}\right)^{3/4}\frac{1}{(250\pi(r+1/r)v_{3})^{\frac{1}{4}}}$ \\
                                         & $\frac{3}{8}\mathcal{A}_{(1)}$ & $\{[5,6,4],1,3\}$ & $\ov{Q}B_{2}Q_{1}$ &\\ \cline{2-4}
                                         & $\frac{1}{24}\mathcal{A}_{(1)}+\frac{1}{12}\mathcal{A}_{(2)}$ & $\{[6,(R,R'),5],[1,3,4],3\}$ & 
$\begin{array}{c} \ov{Q}B_{k}Q_{2} \\ \text{with } k=6,9 \end{array}$ &\\
                                         &                   & $\{[6,(R,R'),5],[1,2,4],3\}$ & $\begin{array}{c} \ov{Q}B_{k}Q_{2} \\ 
\text{with } k=7,10 \end{array}$ &\\
                                         & $\frac{3}{8}\mathcal{A}_{(1)}+\frac{1}{12}\mathcal{A}_{(2)}$ & $\{[5,6,4],[1,3,4],3\}$ & $\ov{Q}B_{3}Q_{2}$ &\\
                                         &                   & $\{[5,6,4],[1,2,4],3\}$  & $\ov{Q}B_{4}Q_{2}$ &\\ \hline
$[(\theta^{2}a),(\theta b'),b']$ & $0$    &  $\{4,1,3\}$        & $\ov{Q}B_{2}Q_{3}$ & $\left(\frac{v_1v_2v_3}{f(S,U)^2}\right)^{3/4}\frac{1}{(250\pi(r+1/r)v_{3})^{\frac{1}{4}}}$ \\ \cline{2-4}
                                               & $\frac{1}{6}\mathcal{A}_{(2)}$ & $\{4,[1,2,1],3\}$ & $\ov{Q}B_{3}Q_{3}$ &\\
                                               & & $\{4,[1,3,1],3\}$ & $\ov{Q}B_{4}Q_{3}$ &\\ \cline{2-4}
                                               & $\frac{1}{12}\mathcal{A}_{(1)}$ & $\{[4,(R,R'),6],1,3\}$ & $\begin{array}{c} \ov{Q}B_{k}Q_{4}
\\\text{with } k=5,8 \end{array}$ &\\ \cline{2-4}
                                               & $\frac{1}{12}\mathcal{A}_{(1)}+\frac{1}{6}\mathcal{A}_{(2)}$ & $\{[4,(R,R'),6],[1,2,1],3\}$ & $\begin{array}{c} \ov{Q}B_{k}Q_{4}
\\\text{with } k=6,9 \end{array}$ &\\
                                               & & $\{[4,(R,R'),6],[1,3,1],3\}$ & $\begin{array}{c} \ov{Q}B_{k}Q_{4}
\\\text{with } k=7,10 \end{array}$ &\\ \cline{2-4}
                                               & $\frac{3}{8}\mathcal{A}_{(1)}$ & $\{[4,(R,R'),4],1,3\}$ & $\begin{array}{c} \ov{Q}B_{k}Q_{3}
\\\text{with } k=5,8 \end{array}$ &\\ \cline{2-4}
                                               & $\frac{3}{8}\mathcal{A}_{(1)}+\frac{1}{6}\mathcal{A}_{(2)}$ & $\{[4,(R,R'),4],[1,2,1],3\}$ & $\begin{array}{c} \ov{Q}B_{k}Q_{3}
\\\text{with } k=6,9 \end{array}$ &\\                                               
                                               & & $\{[4,(R,R'),4],[1,3,1],3\}$ & $\begin{array}{c} \ov{Q}B_{k}Q_{3}
\\\text{with } k=7,10 \end{array}$ &\\
\hline
\end{tabular}
\end{center}
}
\caption{Three-point couplings between the adjoints of $U(3)_a$ and $U(2)_b$ and the quark $Q_i$ and $\ov{Q}$ fields.}
\label{QQbarU2}
\end{table}
%%%%%%%%%%%%%%%%%%

%%%%%%%%%%%%%%%%%%
\begin{table}[ht!]
{\footnotesize
\begin{center}
\setlength{\tabcolsep}{4pt}
\begin{tabular}{|c||c|c|c|c|}
\hline
\muc{5}{|c|}{\text{\bf Couplings of right-handed vector-like quarks to adjoint matter}}
\\\hline\hline
\begin{tabular}{c}sequence\\$[x,y,z]$\\\end{tabular} & \begin{tabular}{c}enclosed\\area\\\end{tabular} & \begin{tabular}{c}triangle\\(or point)\\\end{tabular} 
& coupling & \begin{tabular}{c}K\"ahler factor\\$(K_{xy}K_{yz}K_{zx})^{-1/2}$\\\end{tabular} \\ \hline\hline
$[(\theta a'),a,(\theta a')]$ & 0 & $\{\parallel=[4,5],\parallel,\parallel=[2,3]\}$ & $\begin{array}{c} W_i A_1 W_j \\i=3,5;j=4,6\end{array}$ 
& $\left(\frac{v_1v_2v_3}{f(S,U)^2}\right)^{3/4} \left(\frac{3}{8 \pi^3 r \, v_1v_2v_3}\right)^{1/4}
$
\\\hline
$[a',a,(\theta a)]$ & $\frac{1}{8}\mathcal{A}_{(1)}$ & $\{[4,5,6],1,\parallel=[23]\}$ & $W_{1}A_{2}W_{2}$ & $\left(\frac{v_1v_2v_3}{f(S,U)^2}\right)^{3/4}\frac{1}{(2\pi\,rv_{3})^{\frac{3}{4}}}$ 
\\ \hline
$[(\theta a'),(\theta a),a]$ & 0 & $\{5,1,\parallel=[2,3]\}$ & $\begin{array}{c} W_1 A_2 W_j \\j=4,6\end{array}$ 
& $\left(\frac{v_1v_2v_3}{f(S,U)^2}\right)^{3/4} \left(\frac{\sqrt{3}}{16 \pi^3 r^{2} \, v_2v_3^{2}}\right)^{1/4}
$
\\\hline
$[(\theta a'),a,(\theta^{2}a)]$ & 0 & $\{4,1,\parallel=[2,3]\}$ & $\begin{array}{c} W_i A_2 W_2 \\i=3,5\end{array}$ 
& $\left(\frac{v_1v_2v_3}{f(S,U)^2}\right)^{3/4} \left(\frac{\sqrt{3}}{16 \pi^3 r^{2} \, v_2v_3^{2}}\right)^{1/4}
$
\\\hline\hline
$[a,(\theta^{2}d),(\theta a)]$ & $0$ & $\{5,1,\parallel=[23]\}$ & $Y_{6}Y_{5}A_{2}$ & $\left(\frac{v_1v_2v_3}{f(S,U)^2}\right)^{3/4}\frac{1}{(2\pi\,rv_{3})^{\frac{3}{4}}}$\\ \hline
$[a,d,(\theta^{2}a)]$ & $\frac{1}{16}\mathcal{A}_{(1)}$ & $\{[4,U',6],1,\parallel=[23]\}$ & $Y_{2}Y_{5}A_{2}$ & $\left(\frac{v_1v_2v_3}{f(S,U)^2}\right)^{3/4}\frac{1}{(2\pi\,rv_{3})^{\frac{3}{4}}}$\\ \cline{2-4}
                                     & $\frac{1}{16}\mathcal{A}_{(1)}+\frac{1}{4}\mathcal{A}_{(2)}$ & $\{[4,U',6],[1,1,P],\parallel=[23]\}$ & $Y_{4}Y_{5}A_{2}$ &\\ \hline
$[d,a,(\theta a)]$       & $\frac{1}{16}\mathcal{A}_{(1)}$ & $\{[5,U,6],1,\parallel=[23]\}$ & $Y_{1}A_{2}Y_{6}$ & $\left(\frac{v_1v_2v_3}{f(S,U)^2}\right)^{3/4}\frac{1}{(2\pi\,rv_{3})^{\frac{3}{4}}}$\\ \cline{2-4}
                                     & $\frac{1}{16}\mathcal{A}_{(1)}+\frac{1}{4}\mathcal{A}_{(2)}$ & $\{[5,U,6],[1,1,P],\parallel=[23]\}$  & $Y_{3}A_{2}Y_{6}$ &
\\ \hline
$[d,a,a]$ & $0$ & $\{(U,U'),1,\parallel=[23]\}$ & $Y_{1}A_{1}Y_{2}$ & $\left(\frac{v_1v_2v_3}{f(S,U)^2}\right)^{3/4}\left(\frac{v_{2}}{2\pi^{3}r^{3}v_{1}v_{3}^{3}}\right)^{\frac{1}{4}}$\\ \cline{2-4}
& $0$ & $\{(U,U'),P,\parallel=[23]\}$ & $Y_{3}A_{1}Y_{4}$ &\\ \hline\hline
$[d',a,(\theta a)]$                     & $0$              & $\{5,1,\parallel=[23]\}$              & $Z_{1}A_{2}Z_{6}$ & $\left(\frac{v_1v_2v_3}{f(S,U)^2}\right)^{3/4}\frac{1}{(2\pi\,rv_{3})^{\frac{3}{4}}}$\\ \hline
$[a,(\theta d'),(\theta a)]$       & $\frac{1}{16}\mathcal{A}_{(1)}$ & $\{[5,U,6],1,\parallel=[23]\}$ & $Z_{2}Z_{1}A_{2}$ & $\left(\frac{v_1v_2v_3}{f(S,U)^2}\right)^{3/4}\frac{1}{(2\pi\,rv_{3})^{\frac{3}{4}}}$\\ \cline{2-4}
                                                     & $\frac{1}{16}\mathcal{A}_{(1)}+\frac{1}{4}\mathcal{A}_{(2)}$ & $\{[5,U,6],[1,1,P],\parallel=[23]\}$  & $Z_{4}Z_{1}A_{2}$ &\\ \hline
$[(\theta d'),a,(\theta^{2}a)]$ & $\frac{1}{16}\mathcal{A}_{(1)}$     & $\{[4,U',6],1,\parallel=[23]\}$       & $Z_{3}A_{2}Z_{6}$ & $\left(\frac{v_1v_2v_3}{f(S,U)^2}\right)^{3/4}\frac{1}{(2\pi\,rv_{3})^{\frac{3}{4}}}$\\ \cline{2-4}
                                                     & $\frac{1}{16}\mathcal{A}_{(1)}+\frac{1}{4}\mathcal{A}_{(2)}$ & $\{[4,U',6],[1,1,P],\parallel=[23]\}$ & $Z_{5}A_{2}Z_{6}$ &\\ \hline
$[(\theta d'),a,a]$                     & $0$              & $\{(U,U'),1,\parallel=[23]\}$              & $Z_{3}A_{1}Z_{2}$ & $\left(\frac{v_1v_2v_3}{f(S,U)^2}\right)^{3/4}\left(\frac{v_{2}}{2\pi^{3}r^{3}v_{1}v_{3}^{3}}\right)^{\frac{1}{4}}$ \\ \cline{2-4}
   & $0$              & $\{(U,U'),P,\parallel=[23]\}$              & $Z_{5}A_{1}Z_{4}$ & \\ \hline
\end{tabular}
\end{center}
}
\caption{Three-point couplings between the adjoints of $U(3)_a$ and various types of vector-like quarks with exotic $B-L$ charge.}
\label{tab:qR_to_adjoints}
\end{table}
%%%%%%%%%%%%%%%%%%

%%%%%%%%%%%%%%%%%%%%%%%%%%%%%%%%%%%%%%%%
\begin{table}[ht!]
{\footnotesize
\begin{center}
\setlength{\tabcolsep}{4pt}
\begin{tabular}{|c||c|c|c|c|}
\hline
\muc{5}{|c|}{\text{\bf Couplings of symmetric and adjoint matter of } $U(2)_b$}
\\\hline\hline
\begin{tabular}{c} sequence\\$[x,y,z]$\\\end{tabular} & \begin{tabular}{c}enclosed\\area\\\end{tabular} & \begin{tabular}{c}triangle\\(or point)\\\end{tabular} 
& coupling &  \begin{tabular}{c}K\"ahler factor\\$(K_{xy}K_{yz}K_{zx})^{-1/2}$\\\end{tabular} \\ \hline
$[b,b',(\theta b')]$ & $0$   & $\{(R,R'),i,3\}$       & $b_{i}B_{4+i}\bar{b}_{i}, \; b_{i}B_{7+i}\bar{b}_{i}$  &  $\left(\frac{v_1v_2v_3}{f(S,U)^2}\right)^{3/4}\!\!\!\!\!\!\frac{1}{(64 \pi^2 \sqrt{3} v_1 v_3 (\frac{1}{r} + r))^{\frac{1}{4} }}$\\
                                    &       & $\{(R,R'),i,1\}$       & $b_{3+i}B_{4+i}\bar{b}_{3+i}, \; b_{3+i}B_{7+i}\bar{b}_{3+i} $ & \\
&&& $i=1,2,3$ & \\\cline{2-4} 
                                    & $\frac{1}{6}\mathcal{A}_{(2)}$ & $\{(R,R'),[1,2,3],3\}$  & $\begin{array}{c} b_{i}B_{j}\bar{b}_{k}, \; b_{i}B_{3+j}\bar{b}_{k} \\
\text{with } (ijk)= \\(163),(172),(253),\\(271),(352),(361)\end{array}$ & \\\cline{3-4}
                                    &       & $\{(R,R'),[1,2,3],1\}$ & $\begin{array}{c} b_{i}B_{j}\bar{b}_{k}, \; b_{i}B_{3+j}\bar{b}_{k}\\
\text{with } (ijk)= \\(466), (475), (556),\\ (574), (655), (664)\end{array}$ & \\ \hline
\end{tabular}
\end{center}
}
\caption{Three-point couplings between the symmetrics and adjoints of $U(2)_b$.}
\label{tab:Sym+Adj_U2b}
\end{table}
%%%%%%%%%%%%%%%%%%%%%%%%%%%%%%%%%%%%%%%%%%%

%%%%%%%%%%%%%%%%%%
\begin{table}[ht!]
{\footnotesize
\begin{center}
\setlength{\tabcolsep}{4pt}
\begin{tabular}{|c||c|c|c|c|}
\hline
\muc{5}{|c|}{\text{\bf Three-point couplings of hidden sector fields}}
\\\hline\hline
\begin{tabular}{c}sequence\\$[x,y,z]$\\\end{tabular} & \begin{tabular}{c}enclosed\\area\\\end{tabular} & \begin{tabular}{c}triangle\\(or point)\\\end{tabular} 
& coupling & \begin{tabular}{c}K\"ahler factor\\$(K_{xy}K_{yz}K_{zx})^{-1/2}$\\\end{tabular}\\ \hline
$[(\theta c),h,b]$ & $\frac{1}{16}\mathcal{A}_{(1)}+\frac{1}{8}\mathcal{A}_{(3)}$ & $\{[(T,T'),5,6],1,[4,1,3]\}$ & $\Omega_{2}\Pi^{+}H_{7}$ & $\left(\frac{v_1v_2v_3}{f(S,U)^2}\right)^{3/4}\frac{1}{5(2)^{\frac{1}{4}}}$ \\   \cline{2-4}        
       & $\frac{1}{8}\mathcal{A}_{(1)}+\frac{1}{6}\mathcal{A}_{(2)}+\frac{1}{4}\mathcal{A}_{(3)}$ & $\{[(T,T'),5,6],[1,i,1],[4,1,3]\}$ & $\begin{array}{c}\Omega_{2}\Pi^{+}H_{6+i}\\i\in \{2,3\}\end{array}$ & \\ \hline
$[(\theta h),h,b]$                    & $\frac{1}{16}\mathcal{A}_{(1)}$ & $\{[(T,T'),1,4],1,1\}$ & $X_{2}\Pi^{+}\Pi^{-}$ & $\left(\frac{v_1v_2v_3}{f(S,U)^2}\right)^{3/4}\frac{1}{(250\pi\,rv_{3})^{\frac{1}{4}}}$    \\ \hline
$[(\theta^{2}h),(\theta h),c]$ & $\frac{1}{8}\mathcal{A}_{(1)}$  & $\{[5,1,4],1,4\}$      & $X_{2}\Omega_{1}\Omega_{2}$ & $\left(\frac{v_1v_2v_3}{f(S,U)^2}\right)^{3/4}\frac{1}{(200\pi\,rv_{3})^{\frac{1}{4}}}$ \\ \hline
\end{tabular}
\end{center}
}
\caption{Three-point couplings between the hidden sector fields and Higgses as well as 
 antisymmetrics of $USp(6)_h$ and the $\Pi$ and $\Omega$ fields.}
\label{Hidden-Sequences}
\end{table}

\end{appendix}
%%%%%%%%%%%%%%%%%%%%%%%%%%%

%%%%%%%%%%%%%%%%%%%%
\clearpage
%%%%%%%%%%%%%%%%%%%%%
%%%%%%%%%%%%%%%%%%%%%%%%%%%%%

\addcontentsline{toc}{section}{References}
\bibliographystyle{ieeetr}
\bibliography{refs_Yukawas}

\end{document}